\pdfoutput=1
\documentclass[11pt,twoside,a4paper,cmspaper,final,collab]{cms-tdr}
\def\svnVersion{6950aa4}\def\svnDate{2025/01/15}\def\cmsCernNoTag{CERN-EP-2024-073}\def\cmsCernDate{\today}\def\cmsMessage{Published in Physics Letters B as \href{http://dx.doi.org/10.1016/j.physletb.2024.139088}{\doi{10.1016/j.physletb.2024.139088}.}}
\begin{document}\cmsNoteHeader{HIN-23-001}

\newcommand{\Zjet}{\ensuremath{\PZ{+}\text{jet}}\xspace}
\newcommand{\DIRE} {{\textsc{dire}}\xspace}
\newcommand{\VINCIA} {{\textsc{vincia}}\xspace}
\newcommand{\pp}{\ensuremath{\Pp\Pp}\xspace}
\newcommand{\PbPb}{\ensuremath{\text{PbPb}}\xspace}
\newcommand{\Rg}{\ensuremath{R_{\mathrm{g}}}\xspace}
\newcommand{\xgJ}{\ensuremath{x_{\PGg \mathrm{j}}}\xspace}
\newcommand{\pTjet}{\ensuremath{\pt^{\text{jet}}}\xspace}
\newcommand{\pTg}{\ensuremath{\pt^{\PGg}}\xspace}
\newcommand{\Lres}{\ensuremath{L_{\text{res}}}\xspace}

\newlength\cmsTabSkip\setlength{\cmsTabSkip}{1ex}

\ifthenelse{\boolean{cms@external}}{\providecommand{\cmsLeft}{upper\xspace}}{\providecommand{\cmsLeft}{left\xspace}}
\ifthenelse{\boolean{cms@external}}{\providecommand{\cmsRight}{lower\xspace}}{\providecommand{\cmsRight}{right\xspace}}

\cmsNoteHeader{HIN-23-001}
\title{Girth and groomed radius of jets recoiling against isolated photons in lead-lead and proton-proton collisions at \texorpdfstring{$\sqrtsNN = 5.02\TeV$}{sqrt(s[NN]) = 5.02 TeV}}

\author{The CMS Collaboration}

\date{\today}

\abstract{
    This Letter presents the first measurements of the groomed jet radius \Rg and the jet girth $g$ in events with an isolated photon recoiling against a jet in lead-lead (\PbPb) and proton-proton (\pp) collisions at the LHC at a nucleon-nucleon center-of-mass energy of 5.02\TeV. The observables \Rg and $g$ provide a quantitative measure of how narrow or broad a jet is. The analysis uses \PbPb and \pp data samples with integrated luminosities of $1.7\nbinv$ and $301\pbinv$, respectively, collected with the CMS experiment in 2018 and 2017. Events are required to have a photon with transverse momentum $\pTg > 100\GeV$ and at least one jet back-to-back in azimuth with respect to the photon and with transverse momentum \pTjet such that $\pTjet/\pTg > 0.4$. The measured \Rg and $g$ distributions are unfolded to the particle level, which facilitates the comparison between the \PbPb and \pp results and with theoretical predictions. It is found that jets with $\pTjet/\pTg > 0.8$, \ie, those that closely balance the photon $\pTg$, are narrower in \PbPb than in \pp collisions. Relaxing the selection to include jets with $\pTjet/\pTg > 0.4$ reduces the narrowing of the angular structure of jets in \PbPb relative to the \pp reference. This shows that selection bias effects associated with jet energy loss play an important role in the interpretation of jet substructure measurements.
}

\hypersetup{%
pdfauthor={CMS Collaboration},%
pdftitle={Girth and groomed radius of jets recoiling against isolated photons in lead-lead and proton-proton collisions at sqrt(s[NN]) = 5.02 TeV},%
pdfsubject={CMS},%
pdfkeywords={CMS,  jet substructure, gluons, jets, photon, lead-lead, proton-proton}}

\maketitle 

\section{Introduction}
\label{sec:introduction}

The quark-gluon plasma (QGP) is a high-temperature, strongly interacting, phase of nuclear matter that is produced in ultrarelativistic heavy ion collisions~\cite{Busza:2018rrf,Adams:2012th}. The fragmentation of highly energetic jets, initiated by high-virtuality partons produced in the early stages of the collision, is intertwined with the time evolution of this hot and dense matter. The interactions between the jet and this medium happen over a wide range of energy scales, from the high virtuality of the initial parton of up to hundred \GeV to the medium temperature of hundreds of \MeVns. These interactions are responsible for partonic energy loss in the QGP medium, a phenomenon generally termed as ``jet quenching.'' These interactions are also expected to modify the jet radiation pattern in lead-lead (\PbPb) collisions when compared with proton-proton (\pp) collisions. Jets produced in \pp collisions, which are unaffected by the medium,  are used as a reference for ``vacuum radiation.'' Jets are clustered with a finite radius, and because of the energy redistribution that occurs from the interaction of the jet shower with the medium, there is an effective energy loss with respect to their expected energy if the jet shower had developed in vacuum. The comparison of observables sensitive to medium-induced jet modifications with theoretical model predictions aims to characterize the microscopic structure of the QGP and its abilities to transport matter and energy~\cite{PhysRevD.27.140,Qin:2015srf,Blaizot:2015lma,Casalderrey-Solana:2007knd}. 

The dominant interaction mechanism between the partons created in the parton branching process and the QGP at short distance scales is expected to be medium-induced gluon radiation, leading to a characteristic broadening of the parton shower~\cite{Gyulassy:1990ye,Baier:1996kr,Baier:1996sk}. The QGP medium is characterized by a ``resolution length''~\cite{Casalderrey-Solana:2011ule,Mehtar-Tani:2010ebp}, which sets the minimum angular separation between two partons or subjets that can be resolved by the medium as independent color charges~\cite{Casalderrey-Solana:2012evi}. The additional medium interactions experienced by a jet with larger number of resolved constituents is expected to lead to stronger quenching. The resolution length is directly connected to $\hat{q}$, the transport coefficient of the medium, which corresponds to the average transverse momentum given by the medium to the propagating parton per unit of path length \cite{Baier:1996sk}\cite{Baier:1996kr}\cite{Zakharov:1997uu}. A denser medium (high $\hat{q}$) leads to stronger quenching and to a smaller resolution length. The importance of the medium resolution length in the overall medium-induced modification of the jet has not been conclusively established experimentally, although some models expect it to be the dominant effect.

Another aspect of the jet-medium interaction is that the QGP is found to behave as a strongly-coupled liquid when probed at energy scales that are of the order of its temperature. When the QGP is probed at energy scales much larger than its temperature, the quasiparticle degrees of freedom of the QGP are expected to emerge due to asymptotic freedom~\cite{DEramo:2012uzl,DEramo:2018eoy}. It is expected that jet substructure may give access to these properties.

Experimental results from heavy ion collisions provide evidence for medium-induced jet modifications~\cite{CMS:2011iwn,CMS:2021vui,CMS:2017ehl, Sirunyan:2017bsd, ALICE:2015mdb,ALICE:2019qyj,ALICE:2019ykw,ALICE:2018dxf,ATLAS:2018bvp,ATLAS:2018dgb,ATLAS:2019dsv, STAR:2020ejj, STAR:2021kjt}; recent reviews can be found in Refs.~\cite{Apolinario:2022vzg, Cunqueiro:2021wls}. One of the observables that has been used to study the modifications of the radiation pattern of the jet is the groomed jet radius, \Rg, defined as the rapidity--azimuth ($y$--$\phi$) distance between the two subjets obtained using the soft-drop grooming algorithm~\cite{Butterworth:2008iy, Larkoski:2014wba}, described in Section~\ref{sec:eventselection}. With soft-drop grooming, a jet may be split into two hard subjets that can be used as proxies for the two most energetic partons in the jet shower evolution~\cite{Larkoski:2014uqa}. In \pp collisions, jet grooming is used to remove soft and wide-angle radiation, allowing for a description in the framework of perturbation theory. In \PbPb collisions, in addition to the removal of soft- and wide-angle radiation as in \pp collisions, jet grooming has been proposed as a possible path to study medium-induced jet modifications using properties such as the hardness and broadness of the jet radiation pattern~\cite{Mehtar-Tani:2016aco,Andrews:2018jcm}. With \Rg, we inspect the angular distribution of hard intrajet branchings in search of signatures interpretable in terms of medium-induced effects. These signatures include, in particular, the possible manifestation of the medium resolution length~\cite{Mehtar-Tani:2016aco, Andrews:2018jcm, Caucal:2018dla, Caucal:2021cfb}.

\begin{figure*}[ht]
    \centering
        \includegraphics[width=0.8\textwidth]{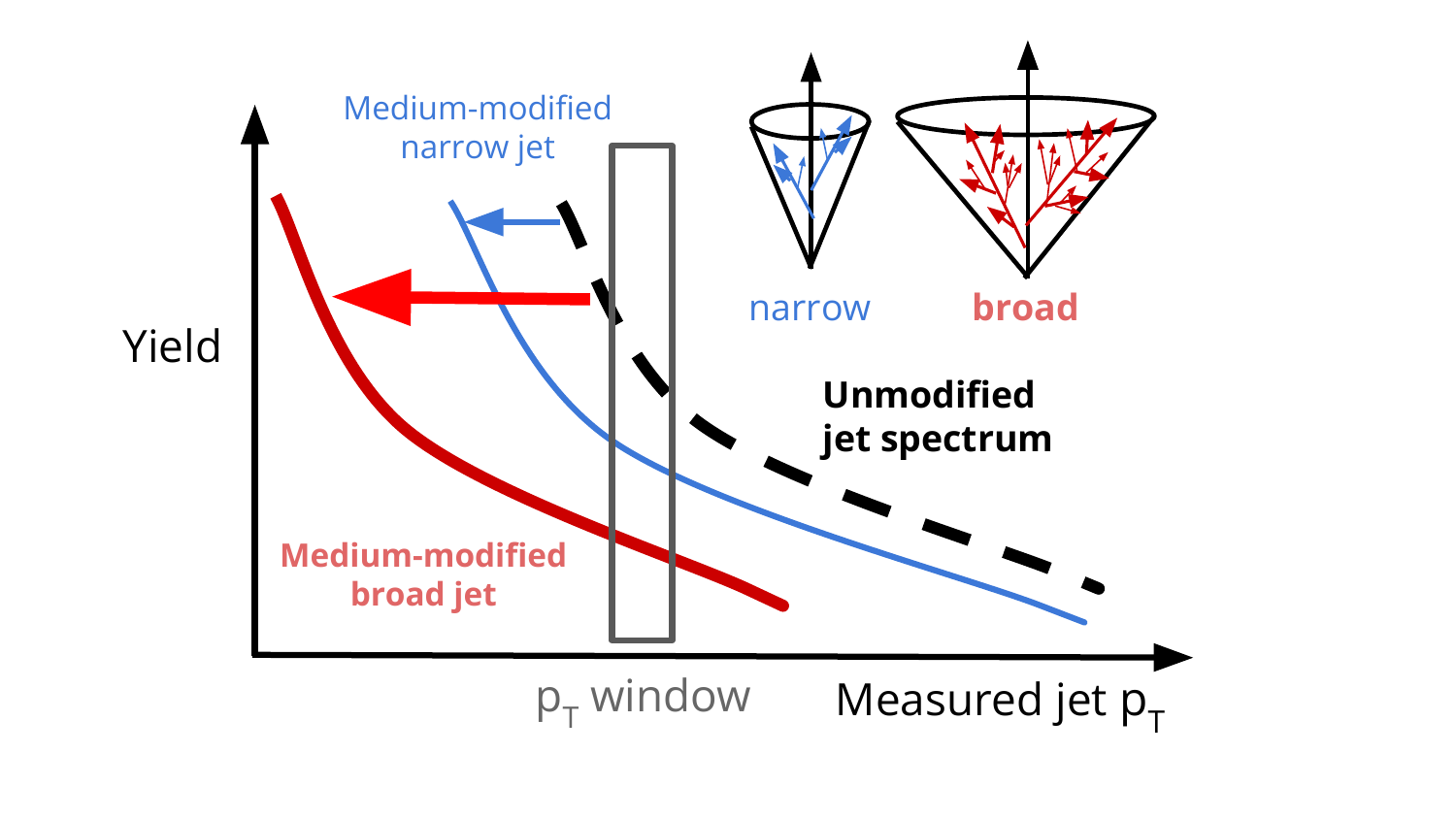}
    \caption{Schematic diagram of the potential selection bias due to jet energy loss that may occur when selecting jets based on the their \pt. Broader structures are expected to be more quenched (thicker red line and arrow), whereas narrower structures are expected to be quenched less (thinner blue line and arrow). Combined with the steeply falling jet \pt spectrum, this can lead to a preferential selection of narrow jets in a given jet \pt interval, as indicated by the vertical rectangular box. The dashed curve represents the jet \pt spectrum in the absence of medium-induced jet modifications.}
    \label{fig:selection_bias}
\end{figure*}

Measurements of \Rg at the RHIC and LHC show a narrower jet substructure in heavy ion collisions compared with \pp collisions for jets reconstructed with the same jet transverse momentum \pt~\cite{ALargeIonColliderExperiment:2021mqf,ATLAS:2022vii, STAR:2020ejj, STAR:2021kjt}. This narrowing of the hard intrajet angular distributions measured via \Rg could be a manifestation of the medium resolution length of the QGP. However, before investigating such a connection, one should assess a potential selection bias that emerges when comparing \PbPb and \pp jets with the same reconstructed \pt~\cite{Brewer:2018dfs,Du:2020pmp,Brewer:2021hmh}. Since jet energy loss may fluctuate jet-by-jet and the jet \pt spectrum is steeply falling, a given jet \pt interval is preferentially populated by jets that are less quenched. The jet shower in heavy ion collisions is expected to factorize into an early vacuum shower, created before the medium formation, followed by a medium-modified showering process~\cite{Caucal:2018dla}. Selecting jets with large \Rg values might preferentially isolate early vacuum radiation patterns that are broad in angle and thus interact more frequently with the QGP, which leads to stronger jet energy loss and causes jets to migrate to lower jet momenta, as illustrated schematically in Fig.~\ref{fig:selection_bias}. Thus, to interpret the modifications of the radiation pattern of the jet in terms of the emergence of other physical scales, such as the medium resolution length, selection bias effects must be understood first.

Prompt photons produced in quark-gluon Compton scattering and quark-antiquark annihilation processes, in which their momentum is balanced by a recoiling jet, provide an ideal topology to study medium-induced jet modifications. A high-momentum photon does not interact strongly with the QGP~\cite{CMS:2020oen,ATLAS:2015rlt,ALICE:2015xmh,STAR:2016use,PHENIX:2005yls}, so its \pt can be used as a proxy for the \pt of the recoiling parton that initiates the jet shower. By selecting on the photon \pt, selection biases associated with jet energy loss should be reduced. Prompt photons can be studied using photon isolation techniques~\cite{CMS:2011nkw,Becher:2022rhu}. Isolated photon+jet events have been used to quantify the jet energy loss via the measurement of the momentum imbalance, defined as the ratio of the \pt of the highest-\pt jet and the photon momentum $\xgJ \equiv \pTjet/\pTg$~\cite{CMS:2017ehl,ATLAS:2018dgb,Sirunyan:2018ncy}. The results show a stronger momentum imbalance in \PbPb collisions than in \pp collisions due to the jet energy loss that occurs in the former. Fragmentation functions have been measured in isolated photon+jet events in \PbPb collisions, showing a suppression of hard jet constituents and an enhancement of soft constituents relative to \pp collisions~\cite{CMS:2018mqn}.

In this Letter, we compare for the first time the groomed jet radius \Rg of jets recoiling against isolated photons, hereafter referred to as ``photon-tagged jets'', in \PbPb and \pp collisions. In addition, we measure the girth $g$ of these jets, described in detail in Section~\ref{sec:eventselection}, which is a \pt-weighted measure of how distant the constituents of the jet are with respect to its axis~\cite{Giele:1997hd, KunnawalkamElayavalli:2017hxo, Wan:2018zpq, ALICE:2018dxf}. Since $g$ does not rely on organizing radiation into clusters of particles, as \Rg does, and since it uses all hadrons in the jet, it potentially highlights different aspects of the jet fragmentation in vacuum and in the medium. Since it uses all particles in the jet, it is sensitive to soft- and wide-angle radiation, in a way such that it complements the observable \Rg. The effective narrowing of the angular structure of jets has also been observed for $g$~\cite{ALICE:2018dxf}. In order to analyze the substructure of strongly quenched jets, but at the same time with a measured jet \pt above the \pt of the underlying event background, the photon transverse momentum, \pTg is selected to be above 100 GeV. The momentum imbalance \xgJ is used as a proxy for jet energy loss in this study. We consider two categories of events: $\xgJ > 0.4$ and $\xgJ > 0.8$. A population of jets with varying degrees of medium modifications is included in the $\xgJ > 0.4$ sample, corresponding to a more inclusive jet selection. The selection with $\xgJ > 0.8$ is less inclusive; it selects jets that experience, on average, less energy loss. A comparison of these two categories is used to demonstrate the effect of a strong selection bias on the substructure of the jet. Tabulated results are provided in the \textsc{HepData} record for this analysis~\cite{hepdata}.

\section{Experimental setup and event reconstruction}
\label{sec:detector}

The analysis uses \PbPb and \pp data collected by the CMS experiment in 2018 and 2017, respectively. The \PbPb and \pp data samples, both at a nucleon-nucleon center-of-mass energy of 5.02\TeV, correspond to integrated luminosities of $1.70\pm 0.03\nbinv$ and $301\pm 6\pbinv$, respectively~\cite{CMS-PAS-LUM-18-002,CMS-PAS-LUM-19-001}.

The CMS apparatus~\cite{CMS:2008xjf,CMS:2023gfb} is a multipurpose, nearly hermetic detector, designed to trigger on~\cite{CMS:2020cmk,CMS:2016ngn} and identify electrons, muons, photons, and (charged and neutral) hadrons~\cite{CMS:2020uim,CMS:2018rym,CMS:2014pgm}. A global ``particle-flow'' (PF) algorithm~\cite{CMS:2017yfk} aims to reconstruct all individual particles in an event, combining information provided by the all-silicon inner tracker and by the crystal electromagnetic (ECAL) and brass-scintillator hadron calorimeters (HCAL), operating inside a 3.8\unit{T} superconducting solenoid, with data from the gas-ionization muon detectors embedded in the flux-return yoke outside the solenoid. The reconstructed particles are used to build \PGt leptons, jets, and missing transverse momentum~\cite{CMS:2018jrd,Khachatryan:2016kdb,CMS:2019ctu}. 

Events of interest are selected using a two-tiered trigger system. The first level (L1), composed of custom hardware processors, uses information from the calorimeters and muon detectors to select events at a rate of around 30\unit{kHz} for \PbPb collisions and 100\unit{kHz} for \pp collisions~\cite{CMS:2020cmk}. The second level, known as the high-level trigger (HLT), consists of a farm of processors running a version of the full event reconstruction software optimized for fast processing, and reduces the event rate to around 1\unit{kHz} before data storage~\cite{CMS:2016ngn}.  

The events are required to have a primary vertex reconstructed within 15\unit{cm} of the nominal interaction point along the beam direction, and within 0.15\unit{cm} in the transverse plane. The additional collisions per bunch crossing have a negligible effect on the measurement. 

The collision centrality in heavy ion collisions is expressed as a percentile of the total inelastic hadronic cross section, following the Glauber methodology~\cite{Loizides:2017ack}, and is determined based on the transverse energy deposited in both hadron forward (HF) calorimeters~\cite{CMS:2011iwn}. The events with close to 0\% centrality, known as ``central'' events, have the largest overlap of the two colliding nuclei, whereas glancing collisions between the two nuclei are referred to as ``peripheral'' events. Hadronic events are selected by requiring at least two towers with an energy larger than 4\GeV in each of the HF calorimeters. In order to focus on medium modification effects, we restrict the analysis to the 30\% most central events, corresponding to \PbPb collisions with relatively small impact parameter.

Jets are clustered from the list of PF candidates using the anti-\kt algorithm~\cite{Cacciari:2008gp} with a distance parameter $R=0.2$. The anti-\kt distance parameter of $R = 0.2$ is a proxy for a circular jet radius of $R = 0.2$ in the $y$--$\phi$ plane. The value of $R = 0.2$, which represents a smaller jet size than the standard $R = 0.4$ value used at the LHC, is chosen to mitigate the sensitivity to uncorrelated background contributions at the level of jets and their substructure~\cite{Andrews:2018jcm}, as discussed in Section~\ref{sec:eventselection}. The four-momentum of the jet is determined using the vector sum of all particle momenta in the jet. For this analysis, jets are required to have $\pTjet > 40\GeV$ and pseudorapidity $\abs{\eta} <2$. In \PbPb collisions, the constituents of the jet are corrected for the underlying event (UE) contribution using the constituent subtraction method~\cite{Berta:2014eza}, an approach that removes or corrects jet constituents based on the average UE density. In \PbPb collisions, the UE comprises the multiple nucleon-nucleon interactions occurring in the same collision. To estimate the average UE density per event, the procedure described in Ref.~\cite{Kodolova:2007hd} was employed using PF candidates. The mean value and dispersion of the transverse energies from the PF candidates are calculated in several $\eta$ bins for each event, as done in Refs.~\cite{CMS:2011iwn, CMS:2012ulu, CMS:2017ehl, Sirunyan:2017bsd, CMS:2021vui}. Due to region-to-region fluctuations of the UE, residual background fluctuations smear the jet \pt and its internal structure~\cite{ALICE:2012nbx}, which are corrected with unfolding, as described in Section~\ref{sec:unfolding}. The presence of fake prongs from the combinatorial background at the substructure level would manifest as an additional bump-like structure at large \Rg values~\cite{Andrews:2018jcm}. With the $z_\text{cut}=0.2$ and $R = 0.2$ requirements, no such structure is observed at the detector level. To account for possible data-to-simulation differences in the modeling of the UE by \HYDJET, a dedicated uncertainty is estimated (labeled ``Centrality'' in Section \ref{sec:unfolding}). The jets are corrected for the detector response with jet energy corrections derived from independent \PbPb and \pp simulations along with additional corrections for the imperfect modeling of the detector response~\cite{CMS:2011shu} from dijet and photon+jet studies using high-luminosity \pp data from the same LHC running period. In addition to the jet energy correction, the momentum resolution of jets is larger in data compared with simulation.  Corrections to the simulation to account for the momentum resolution differences are derived from dijet balancing studies performed in \pp collisions at $\sqrt{s} = 13\TeV$ in 2017 and 2018~\cite{Khachatryan:2016kdb}. To incorporate this effect, a Gaussian smearing is applied to the detector-level jet \pt values in simulation to match the resolution in data.  This smeared mapping is then applied in the unfolding procedure, as described in Section~\ref{sec:unfolding}. 

Photons in \PbPb and \pp collisions are reconstructed using the algorithms described in Ref.~\cite{CMS:2020uim}. For this analysis, photons are required to have $\pTg > 100\GeV$ and $\abs{\eta_{\PGg}} < 1.44$ where the photon triggers are fully efficient in both \PbPb and \pp systems. Reconstruction, identification, and energy correction algorithms have been optimized to perform in the extreme conditions of high UE activity in central \PbPb collisions~\cite{CMS:2020uim}.

The \PYTHIA{ 8.230}~\cite{Sjostrand:2014zea} event generator with tune CP5~\cite{CMS:2019csb} is used to calculate Monte Carlo (MC) corrections and for making comparisons with the data. For \pp simulations, a second sample is generated at leading-order (LO) by \HERWIG{ 7.2.2}~\cite{Gieseke:2003rz, Ball:2014uwa,Webber:1983if} with the CH3 tune~\cite{Sirunyan:2020pqv} to assess systematic uncertainties related to the modeling of the parton shower, hadronization, multiparton interactions, and beam-beam remnant interactions. All generated samples are passed through a detailed simulation of the CMS detector using \GEANTfour~\cite{Agostinelli:2002hh}. In \pp collisions, the UE consists of multiparton interactions and those of beam-beam remnants. In \PbPb collisions, the UE encompasses, in addition, the multiple nucleon-nucleon interactions occurring in the same \PbPb collision in addition to the hard scattering. Thus, for the \PbPb simulation, \PYTHIA{8} CP5 events are embedded into a heavy ion UE produced with \HYDJET{ v1.9}~\cite{Lokhtin:2005px}, which is tuned to reproduce the global event properties of \PbPb collisions, including the mean average density and its fluctuations. This was verified with random cone studies using $R=0.2$ cone jets.

\section{Analysis method}
\label{sec:eventselection}

The groomed jet radius \Rg is the $y$--$\phi$ distance between the two subjets found with the soft-drop grooming algorithm. This algorithm consists of first reclustering the jet constituents obtained from the anti-\kt jet~\cite{Cacciari:2008gp} using the Cambridge--Aachen algorithm, which is a pairwise clustering algorithm that clusters particles (and subjets thereafter) that are closer in rapidity and azimuth, imposing angular ordering~\cite{Dokshitzer:1997in, Wobisch:1998wt}. Then, the Cambridge--Aachen clustering history is undone iteratively, always undoing the hardest subjet at each step of the iteration, until the first pair of subjets that satisfies the condition $z > z_\text{cut} \theta ^{\beta}$ is found. Here, $z$ corresponds to the momentum fraction $z=\pt^\mathrm{sub} /(\pt^\mathrm{sub}+\pt^\mathrm{lead})$ where $\pt^\mathrm{lead}$ ($\pt^\mathrm{sub}$) is the momentum of the harder (softer) subjet in the declustering step and $\theta$ the distance in $y$--$\phi$ between the harder and softer subjets in the declustering procedure, which corresponds to \Rg for the pair of subjets obtained by the soft-drop algorithm. We use the parameters $z_\text{cut}=0.2$ and $\beta=0$ for soft-drop grooming, which allows us to better control the large UE background of \PbPb collisions~\cite{Mulligan:2020tim}. These soft-drop grooming parameter values were also used by the ALICE and ATLAS Collaborations~\cite{ALargeIonColliderExperiment:2021mqf,ATLAS:2022vii}. All the PF constituents of the original anti-\kt jet are used in the Cambridge--Aachen clustering without an explicit selection requirement on their \pt. The PF candidates used for the determination of the observables are the constituent-subtracted PF candidates.

In addition to \Rg, we measure the jet girth $g$~\cite{Giele:1997hd, KunnawalkamElayavalli:2017hxo, Wan:2018zpq, ALICE:2018dxf,ALICE:2021njq,CMS:2021iwu}, which is defined as the sum of the product of the momentum fraction of the jet constituents and their distance relative to the anti-\kt jet axis, namely:
\begin{equation}
g= \frac{1}{{\pTjet}} \sum_{i }  \pt^{i} \Delta R_{i,\text{jet}} ,
\end{equation}
where $\Delta R_{i,\text{jet}} = \sqrt{\smash[b]{(\Delta y_{i, \text{jet}})^2+(\Delta \varphi_{i, \text{jet}})^2} }$ is the distance in $y$--$\phi$ of the $i$-th jet constituent with respect to the anti-\kt jet axis.
For the definition of the observables, rapidity $y$ is used at both detector and particle levels. For other matters, such as the calibration of jets and photons, pseudorapidity $\eta$ is a more appropriate variable in order to account for detector geometry dependence, so such objects are selected based on $\eta$.

This analysis requires the presence of at least one isolated photon with $\pTg > 100\GeV$ and $\abs{\eta_{\PGg}} < 1.44$ and the presence of at least one anti-\kt jet with $R=0.2$ and $\abs{\eta_\text{jet}} < 2$ carrying at least 40 or 80\% of \pTg ($\xgJ > 0.4$ or 0.8). The highest-\pt isolated photon is used in the analysis. The $\xgJ > 0.4$ requirement is applied so that jets have a minimum \pt of 40\GeV, whereas the $\xgJ > 0.8$ selection is motivated by the \pt shift caused by out-of-cone radiation for jets produced in \pp collisions~\cite{Dasgupta:2007wa}.  The isolated photon and the jets have a separation in azimuthal angle defined by $\Delta \varphi_{\PGg j} > \frac{2}{3}\pi$. The selection requirements are such that the reconstructed jet primarily originates from the hard scattering, and reduces the likelihood that the jet originates from the large UE fluctuations of more central \PbPb collisions. 
The shapes of the normalized $\Delta \phi_{\PGg \mathrm{j}}$ distributions are compatible, within the statistical uncertainties, between \PbPb and \pp, suggesting that the contribution of fake jets is negligible within the precision of the measurement. The presence of fake jets would manifest as an additional flat, $\Delta \phi_{\PGg \mathrm{j}}$-independent contribution from pairings of isolated photons with fake jets from the UE.
In turn, the negligible residual combinatorial background enables the application of unfolding corrections to the particle level that are robust with respect to detector smearing and the large UE background effects on jet \pt. Overall, the small radius $R=0.2$ of the jet and the \pt requirements of the photon and jet represent a balance between managing the large UE activity and preserving our ability to reach lower values of \xgJ using hard jets.

The UE in \PbPb collisions can create spurious structures at the level of the jet substructure. This is mitigated with the use of a small distance parameter $R = 0.2$, since the contribution of particles from the UE scales with $R^2$~\cite{Andrews:2018jcm}. In addition, soft-drop grooming with $z_\text{cut} = 0.2$ is used to further suppress these contributions~\cite{Andrews:2018jcm}. The reason is that subjets created by the large UE tend to have a more asymmetric momentum balance compared with the expectation from parton branchings.

For parton energy loss studies in heavy ion collisions, a selection of prompt photons is desired since they provide a more reliable proxy for the momentum of the recoil parton that initiated the jet shower. To suppress the contribution from nonprompt photons, a cut-based approach is followed, where the selection is optimized using binary trees to maximize the signal significance squared~\cite{TMVA:2007ngy}.  The prompt photon signal, which includes both direct and fragmentation photon contributions, is identified based on criteria derived from MC simulated events. We require the scalar \pt sum of all the particles around the photon within a radius of 0.4 in units of $\eta$--$\phi$ to be less than 5 \GeV at particle level. Residual inefficiencies for the signal are accounted for in the corrections described in Section~\ref{sec:unfolding}. The detector-level variables used in the optimization of the photon selection are the fraction of hadronic energy around the photon candidate $H/E$, the shower shape variable $\sigma_{\eta\eta}$, and the isolation variable $\mathcal{I}$~\cite{CMS:2020uim}. The fraction $H/E$ is calculated as the ratio of HCAL over ECAL energy within an $\eta$--$\phi$ distance of 0.15 units with respect to the photon candidate. The $\sigma_{\eta\eta}$ variable quantifies the lateral energy spread in the ECAL cluster, which is typically wider for photons from neutral-hadron decays than for prompt photons. The isolation variable $\mathcal{I}$ is given by the sum of transverse energies in ECAL and HCAL and the transverse momenta of all tracks with $\pt > 2\GeV$ within an $\eta$--$\phi$ distance of 0.4 with respect to the photon candidate, which is corrected by subtracting the estimated average energy of the UE~\cite{CMS:2020oen}. As described in Ref.~\cite{CMS:2020oen}, a rectangular area in pseudorapidity and azimuthal angle is used to estimate the average energy density.’ The $\mathcal{I}$ variable can have negative values as a result of the aforementioned UE subtraction. In \PbPb events, we select isolated photons with $\mathcal{I} < 2.1\GeV$ while in \pp events the requirement is $\mathcal{I} < -0.1\GeV$, which corresponds to a particle level $\mathcal{I}$ value of less than 5\GeV for generated photons. The different selection requirements in the \pp and PbPb samples are optimized independently such that the identification efficiency for isolated photons is similar for the two systems.

The direct photon purity of the cut-based selected sample is applied as a correction for any photons from hadronic decay or fragmentation. The contribution of photon-tagged jets originating from the decays of neutral hadrons is taken into account via signal photon purity of the cut-based selected sample. The photon purity estimation is obtained by performing a template fit of the shower shape distribution where the signal template is MC-based and the background template is built with nonisolated ($10 < \mathcal{I} < 20\GeV$) photons in the data~\cite{CMS:2020oen}. The region $10 < \mathcal{I} < 20\GeV$ is depleted in signal contributions from prompt photons according to simulation studies.

In order to estimate the uncertainty on the photon purity, an alternative matrix ("ABCD") method is used relying on data and is independent of the template fit method. The ABCD method consists of dividing the photon+jet data sample into four mutually exclusive regions using a two-dimensional plane with the variables $\sigma_{\eta\eta}$ and $\mathcal{I}$. We assume that the selection efficiencies of $\sigma_{\eta\eta}$ are independent of $\mathcal{I}$ for the background, which is supported by simulation studies. Three of the four regions are dominated by the background (the B, C and D regions), while the fourth region (A) is a mixture of signal and background. The samples A, B, C, and D are comprised of photon-tagged jets whose photons satisfy the four cut combinations of the selections on $\sigma_{\eta\eta}$ and $\mathcal{I}$ used in the template fit. The expected yield of background events in the signal dominated sample A is estimated using the ratio of yields in the C and D samples and extrapolating it to the regions A and B following a procedure similar to Ref.~\cite{ALICE:2019rtd,CMS:2021klw}. According to simulation studies, it is expected that the amount of remaining prompt photon events in the B, C, D regions is negligible with respect to the background. The signal inefficiency from the ABCD or template-fit methods, which is at the per mille-level, is taken into account as part of the corrections described in Section~\ref{sec:unfolding}. The estimated photon purities have values of $0.77 \pm 0.01$ and $0.93 \pm 0.02$ for \PbPb and \pp collisions, respectively, where the central value corresponds to the photon purity obtained with the template fit method and the uncertainties correspond to the symmetrized difference of the photon purity obtained with the alternative ABCD method.

The number of reconstructed events with $\xgJ > 0.4$ (0.8) is 4717 (1940) in \PbPb collisions, and is 20636 (10796) in \pp collisions. According to simulations using LO matrix elements for the hard scattering, the sample of photons used in this analysis is dominated by prompt photons from quark-gluon Compton scattering and quark-antiquark annihilation processes. The contribution from Bremsstrahlung photons is about 15\%, primarily in the $\xgJ > 1$ region~\cite{dEnterria:2012kvo}.

\section{Unfolding and systematic uncertainties}
\label{sec:unfolding}

To facilitate comparisons with theoretical predictions and with other experiments, we unfold the detector-level distributions to the level of stable particles. Two unfolding corrections are performed: one set for (\xgJ, \Rg) and the other for $(\xgJ, g)$. The corrections to the particle level account for bin-to-bin migration effects, underlying event background, and acceptance and efficiency corrections. To quantitatively describe the migration effects, a multidimensional response matrix is calculated using isolated photon+jet events at the particle and detector levels. Both isolated photons and recoiling jets at the particle and detector levels are matched by proximity in $\eta$--$\phi$ space. The migrations in \xgJ are dominated by the smearing of the jet \pt. The unfolded distributions are normalized to the number of jets whose substructure is analyzed and that pass the given \xgJ selection, $N_\text{jet}$. In the case of soft-drop groomed jets, this includes the jets that fail the soft drop condition. This choice of normalization provides sensitivity not only to the shape of the distribution but also to the yield of splittings passing the SD condition in the given jet \pt range. Photon losses due to the photon \pt resolution effects are negligible for the observables reported in the measurement, which are normalized per photon+jet pair. It was verified that the unfolded solution is stable at the chosen number of iterations, within the statistical uncertainties. This stability test consists of comparing the unfolded solution at different subsequent iterations relative to the chosen iteration.

Due to the sensitivity to statistical fluctuations in the detector-level distribution, we use regularized unfolding. We use D'Agostini iterative unfolding with early stopping~\cite{DAgostini:1994fjx}, as implemented in the \textsc{RooUnfold} package~\cite{RooUnfold}. Unfolding corrects the full jet \pt (via \xgJ) and the jet substructure variables. Bin-by-bin matching purity and efficiency corrections are applied before and after the unfolding, respectively. The matching purity correction accounts for the fraction of detector-level jets that are not assigned to a generator level jet, whereas the efficiency correction accounts for the fraction of particle level jets that are not associated with a detector-level jet. The matching purity and efficiency corrections are of a few percent and nearly independent of \Rg and $g$. The binning choice ensures a sufficiently large number of counts per bin for stable unfolding corrections. The number of iterations, which plays the role of the regularization parameter in D'Agostini unfolding, is such that the unfolded solution folded back to detector level is statistically compatible with the input distribution, the measured one. The number of iterations for \PbPb is seven and nine for \Rg and $g$ in the $\xgJ > 0.4$ category. It is slightly lower for $\xgJ > 0.8$ in \PbPb, which is three and six iterations for \Rg and $g$, respectively. Fewer iterations are needed in pp in general, four and five (three and two) iterations for \Rg and $g$ for the $\xgJ > 0.4$ ($\xgJ > 0.8$) category. 

Theoretical and experimental uncertainties are propagated in the unfolding procedure through variations of the response matrix, as well as the purity, and efficiency corrections. The following sources of systematic uncertainties are considered:

\textit{Physics model dependence:} 
Regularized unfolding introduces a bias towards the input or prior MC generator-level spectrum. To assess this bias, we use other assumptions for the prior spectrum to cover a reasonable range of possibilities. In addition, this results in a change in the detector migration matrix itself that is used for the unfolding corrections.

In \pp collisions, the nominal MC sample used for unfolding is \PYTHIA{8} with the CP5 tune. We use \HERWIG{7.1.4}~\cite{Bellm:2015jjp,Bellm:2019zci} simulated events as an alternative MC sample, using the CH3 tune~\cite{Sirunyan:2020pqv}. The \HERWIG{7} generator has an angular-ordered shower to account for color coherence effects, distinct from the \pt-ordered shower of \PYTHIA{8}. \HERWIG{7} uses the cluster fragmentation model, different from the string model used in \PYTHIA{8}. We take the difference of the unfolding determined with these variations relative to the nominal unfolded results as the respective systematic uncertainty. The resulting uncertainty is symmetrized bin by bin. 

For \PbPb collisions, we considered a variation of the \PYTHIA{8} CP5 events with a modification of its quark-gluon jet fraction, which is used as a proxy of medium-modification effects. It is expected that medium-induced jet modifications in general leads to a larger number of particles and momentum broadening. One way of assessing this effect in the corrections is by increasing the fraction of gluon jets. Quark and gluon jets have different shapes in \Rg and $g$; quark jets tend to be narrower than gluon jets on average. The modification of the quark-gluon jet fraction in simulation is done by fitting the jet substructure observables at the detector level using a template fit with quark and gluon jet templates from simulated \PYTHIA{8} CP5 events. The template fitting yields a sample of MC events with a larger fraction of gluon-initiated jets (from 45\% to about 65\%) for $\xgJ > 0.4$ while reducing the fraction of gluon-initiated jets (from 25\% to about 12\%) for $\xgJ > 0.8$ compared with the generator level yields, which we use to reweigh the nominal \PYTHIA{8} CP5 sample. This modification of the quark-to-gluon jet fraction is one way of modifying the composition of broad and narrow jets in simulation. Jet quenching is expected to lead to modifications of the hardness and width of jet radiation patterns, so this is one way of mimicking such an effect. Because the quark and gluon jets used in the simulation are derived for vacuum showers of \PYTHIA{8}, a physical interpretation of the changes of the quark-gluon jet fraction is not drawn for these changes. The unfolding procedure is repeated with this variation of the \PYTHIA{8} CP5 sample using reweighted events according to the quark/gluon jet fraction found in the template fit. The symmetrized difference with respect to the nominal result is used as the respective systematic uncertainty.

\textit{Regularization bias uncertainty:} To quantify the uncertainty associated with the choice of the number of iterations in D'Agostini unfolding, we consider the difference between the nominal unfolding solution and the unfolding solution obtained with a different number of iterations. The response matrix and prior spectrum are fixed for the evaluation of this uncertainty. The alternative number of iterations corresponds to the optimal number of iterations obtained when unfolding with the alternative MC generator used to estimate the model uncertainty.

\textit{PF candidate energy scale uncertainty:} To assess the impact of the individual energy calibration uncertainties of the PF candidates that are used for the jet substructure, we scale the four-momenta of the charged hadron and photon PF candidates by $\pm1$\% and of neutral hadron PF candidates by $\pm3$\% at the detector level in simulation~\cite{CMS:2021iwu}. The PF energy scale variations are done in an uncorrelated way, \ie, an up or down shift in the four-momenta for each PF candidate species at a time, such that six different variations of PF energy scale shifts are evaluated. The unfolding corrections are repeated for each variation, and the difference with respect to the nominal unfolding result is evaluated. This has a different effect in \Rg and $g$. For \Rg, it changes the fraction of subjets that previously failed (or passed) the soft-drop grooming condition, and it mildly modifies the substructure distribution itself via smearing of the subjet \pt. For $g$, since it is defined as the sum of the products of \pt of the constituents and their $y$--$\phi$ distance to the jet axis, the shift has a direct effect on its shape and thus the effect is stronger than for \Rg.

\textit{Jet energy scale (JES) uncertainties:} The uncertainty in the JES is evaluated from dijet and \GAMJET \pt balancing methods~\cite{Khachatryan:2016kdb}. The uncertainty in the JES in \pp collisions is around 3--4\%, increasing as a function of $\abs{\eta}$.  In \PbPb collisions, the uncertainty is around 4\%, except for the barrel-endcap transition region ($1.3 < \abs{\eta} < 1.6$), where it can become as large as 10\%. An additional source of uncertainty in the JES is considered in \PbPb collisions to take into account the differences in the particle mixture in simulation and data due to jet energy loss~\cite{CMS:2017ehl}. We construct an alternative set of unfolding corrections by varying the JES within its uncertainties. The JES uncertainty does not affect the jet substructure itself, since it is related to the calibration of the full jet. However, it can lead to migration effects via the $\xgJ > 0.4$ (0.8) threshold used in our selection. 

\textit{Jet energy resolution uncertainties:} The uncertainty in the jet energy resolution is evaluated from a dijet balancing method~\cite{Khachatryan:2016kdb}. In \pp collisions, the uncertainty in the resolution is in the range of 2--4\% in the barrel region, but is larger in the endcap and transition regions, where it varies in the range of 10--20\%, depending on $\eta$. To propagate the uncertainty from jet \pt resolution, the data-to-simulation resolution scale factors are varied according to their uncertainties and the unfolding corrections are repeated. In \PbPb collisions, there is an additional contribution to the jet energy resolution uncertainty because of the modeling of the UE in simulation with \HYDJET. This uncertainty is evaluated by comparing the energy in randomly-distributed cones in data and simulation. The difference in data and simulation on the energy distributions obtained using the random-cone method is used to estimate the effect on the jet resolution. To estimate this uncertainty, we shift the centrality interval in simulation. To account for these data-to-simulation differences, in practice the centrality in simulation is shifted by 4.5\% for the nominal results, i.e, 0--30\% central events in data correspond to 4.5--34.5\% in simulation. For the respective uncertainty, we consider the intervals of 3--33\% and 6--36\% for the down and up variations, respectively. 

\textit{Photon background subtraction uncertainty:} Nominally, the normalization of the photon background from neutral hadron decays is determined using a template fit of the $\sigma_{\eta\eta}$ distribution using simulated signal photon+jet events and the shower shape distribution in data in a region dominated by background (sideband in $\mathcal{I}$). As an alternative method, we calculate the normalization of the background using control samples in data via the ABCD method, as explained in Section~\ref{sec:eventselection}. The unfolding is repeated after this neutral-hadron photon decay background subtraction variation and the corresponding difference with respect to the yields extracted with the template fit is considered as the systematic uncertainty, which is symmetrized.

\textit{Response matrix statistical uncertainties:} We consider the statistical uncertainties in the MC sample used to construct the response matrix as an additional systematic uncertainty, which is propagated in the unfolding procedure at each iteration. 

The uncertainties are added in quadrature bin-by-bin for the final combination of systematic uncertainties. The uncertainties are considered to be bin-to-bin fully correlated, except for the statistical uncertainties of the unfolded distributions and the response matrix statistical uncertainties, whose covariance matrices are determined directly as part of the unfolding procedure \textsc{RooUnfold}. The relative bin-by-bin uncertainties are shown in Table~\ref{tab:systematics_xJgp4} for $\xgJ > 0.4$ and in Table~\ref{tab:systematics_xJgp8} for $\xgJ > 0.8$ in both \pp and \PbPb collisions. The dominant contribution to the measurement uncertainty is the model dependence used in the corrections, with the PF energy scale uncertainty being the subleading one. The uncertainties grow at large values of $g$ and \Rg. Statistical uncertainties are commensurate with the systematic uncertainties for the \PbPb measurement, whereas they are smaller than the systematic ones for \pp. Since the radiation pattern is modified in nontrivial ways in the case of \PbPb collisions and since the \PbPb and \pp data were collected in different years, we consider the systematic uncertainties to be uncorrelated between the \PbPb and \pp measurements.

\begin{table*}[hbtp!]
    \centering
    \topcaption{Summary of bin-by-bin percentual relative uncertainties for $\xgJ > 0.4$.}
    \label{tab:systematics_xJgp4}
    \begin{tabular}{lccccc}
        \hline
        Uncertainty source                  & \multicolumn{2}{c}{$g$} & \multicolumn{2}{c}{\Rg}            \\
                                            &           \pp       &             \PbPb      &               \pp     &           \PbPb          \\
        \hline
        Physics model dependence            &           1.3--7.5   &             1.2--2.5   &             0.2--5.3   &             0.9--5.4     \\
        Regularization bias                 &           0.1--0.7   &   $\lesssim$0.1--1.2   &        $\lesssim$0.1   &        $\lesssim$0.1     \\
        Photon PF energy scale              &           0.4--1.5   &             1.8--5.1   &             0.2--0.6   &             0.1--2.9     \\
        Charged hadron PF energy scale      &           0.6--4.1   &             0.5--5.3   &   $\lesssim$0.1--0.5   &   $\lesssim$0.1--1.6     \\
        Neutral hadron PF energy scale      &           0.1--1.5   &             0.2--5.3   &             0.1--0.5   &             0.3--2.4     \\
        JES                                 &           0.2--3.3   &             0.2--2.6   &             0.1--1.5   &             0.3--3.3     \\
        JER                                 & $\lesssim$0.1--2.0   &   $\lesssim$0.1--3.7   &   $\lesssim$0.1--0.2   &             0.1--1.8     \\
        Centrality                          &             \NA      &             0.6--4.0   &               \NA      &   $\lesssim$0.1--2.4     \\
        Photon background subtraction       &           0.1--0.3   &        $\lesssim$0.1   &   $\lesssim$0.1--0.2   &        $\lesssim$0.1     \\
        Response matrix statistical         &           1.0--2.9   &             1.4--4.5   &             0.9--2.2   &             1.4--3.6     \\[\cmsTabSkip]
        Total systematic                    &           2.2--9.8   &             2.7--10.7  &             1.3--6.0   &             2.7--8.5     \\[\cmsTabSkip]
        Total statistical                   &           1.4--3.5   &             3.5--7.6   &             1.4--2.5   &             3.6--6.4     \\
        \hline
    \end{tabular}
\end{table*} 

\begin{table*}[hbtp!]
    \centering
    \topcaption{Summary of bin-by-bin percentual relative uncertainties for $\xgJ > 0.8$.}
    \label{tab:systematics_xJgp8}
    \begin{tabular}{lccccc}
        \hline
        Uncertainty source                  & \multicolumn{2}{c}{$g$} & \multicolumn{2}{c}{\Rg}            \\
                                            &           \pp       &             \PbPb    &               \pp     &           \PbPb        \\
        \hline
        Physics model dependence            &           0.3--8.9   &           0.1--7.3   &             0.6--3.0   &           0.1--5.7     \\
        Regularization bias                 &           0.3--0.9   &           0.1--2.0   &   $\lesssim$0.1--1.6   &           0.6--2.9     \\
        Photon PF energy scale              &           0.1--1.3   &           0.1--4.4   &             0.1--1.4   &           0.2--0.9     \\
        Charged hadron PF energy scale      &           0.5--3.7   &           0.1--8.6   &   $\lesssim$0.1--2.5   &           0.1--1.3     \\
        Neutral hadron PF energy scale      &           0.0--2.4   &           0.2--7.5   &             0.1--1.9   &           0.1--3.7     \\
        JES                                 &           0.7--4.8   &           1.5--7.8   &             0.0--5.3   &           2.0--6.8     \\
        JER                                 & $\lesssim$0.1--1.0   &           0.2--3.0   &   $\lesssim$0.1--2.1   &           0.3--3.0     \\
        Centrality                          &             \NA      &           0.4--5.9   &               \NA      &           0.1--2.5     \\
        Photon background subtraction       & $\lesssim$0.1--0.1   & $\lesssim$0.1--0.1   &             0.1--0.1   & $\lesssim$0.1--0.1     \\
        Response matrix statistical         &           1.0--3.8   &           1.5--5.9   &             1.0--3.2   &           1.1--4.0     \\[\cmsTabSkip]
        Total systematic                    &           1.6--11.7  &           2.5--14.8  &             2.7--7.5   &           2.6--10.6    \\[\cmsTabSkip]
        Total statistical                   &           1.4--5.1   &           4.2--14.6  &             1.6--3.9   &           3.6--11.6    \\
        \hline
    \end{tabular}
\end{table*} 

\section{Results}
\label{sec:results}

\begin{figure}[htbp!]
    \centering
        \includegraphics[width=0.49\textwidth]{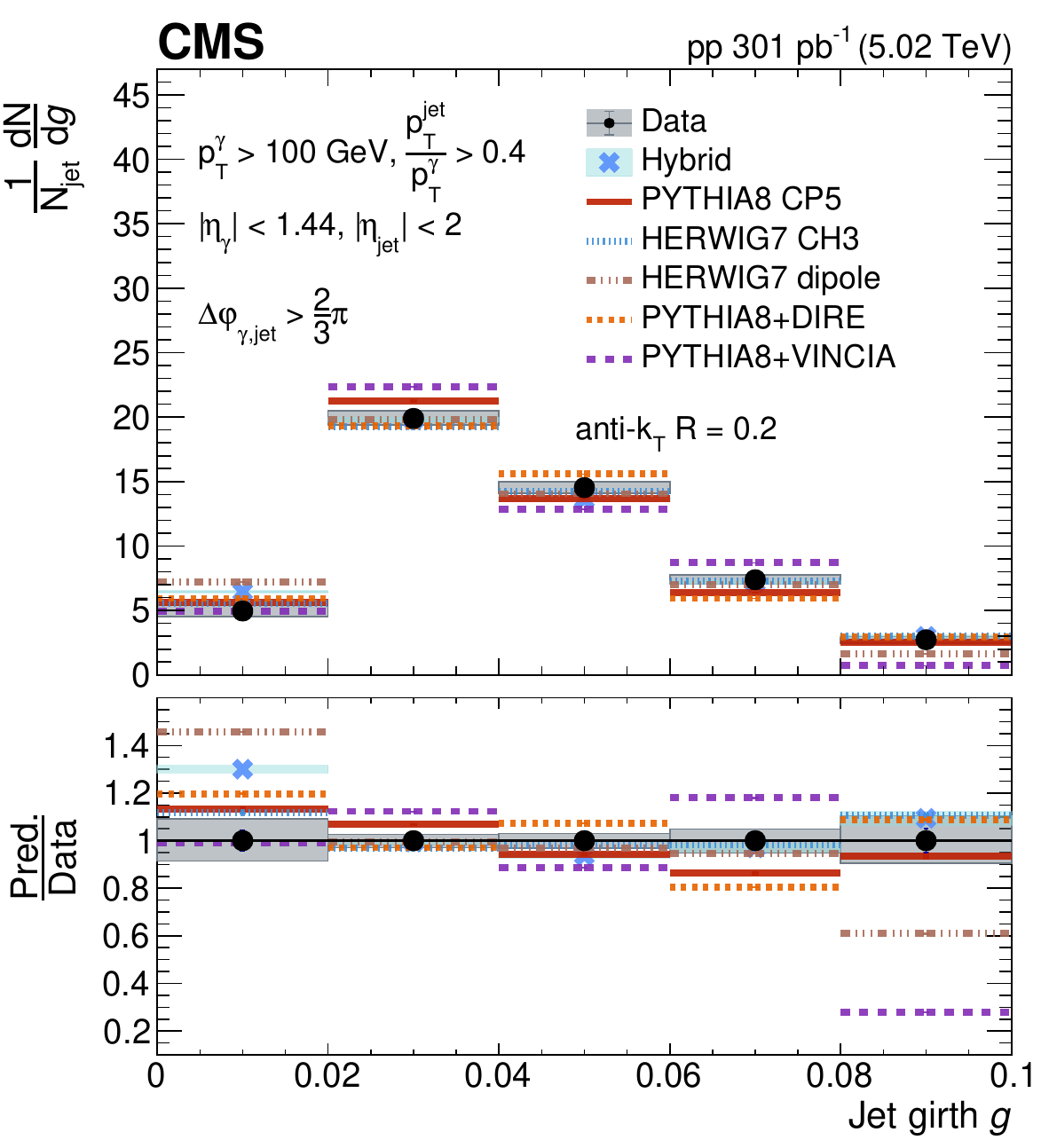}
        \includegraphics[width=0.49\textwidth]{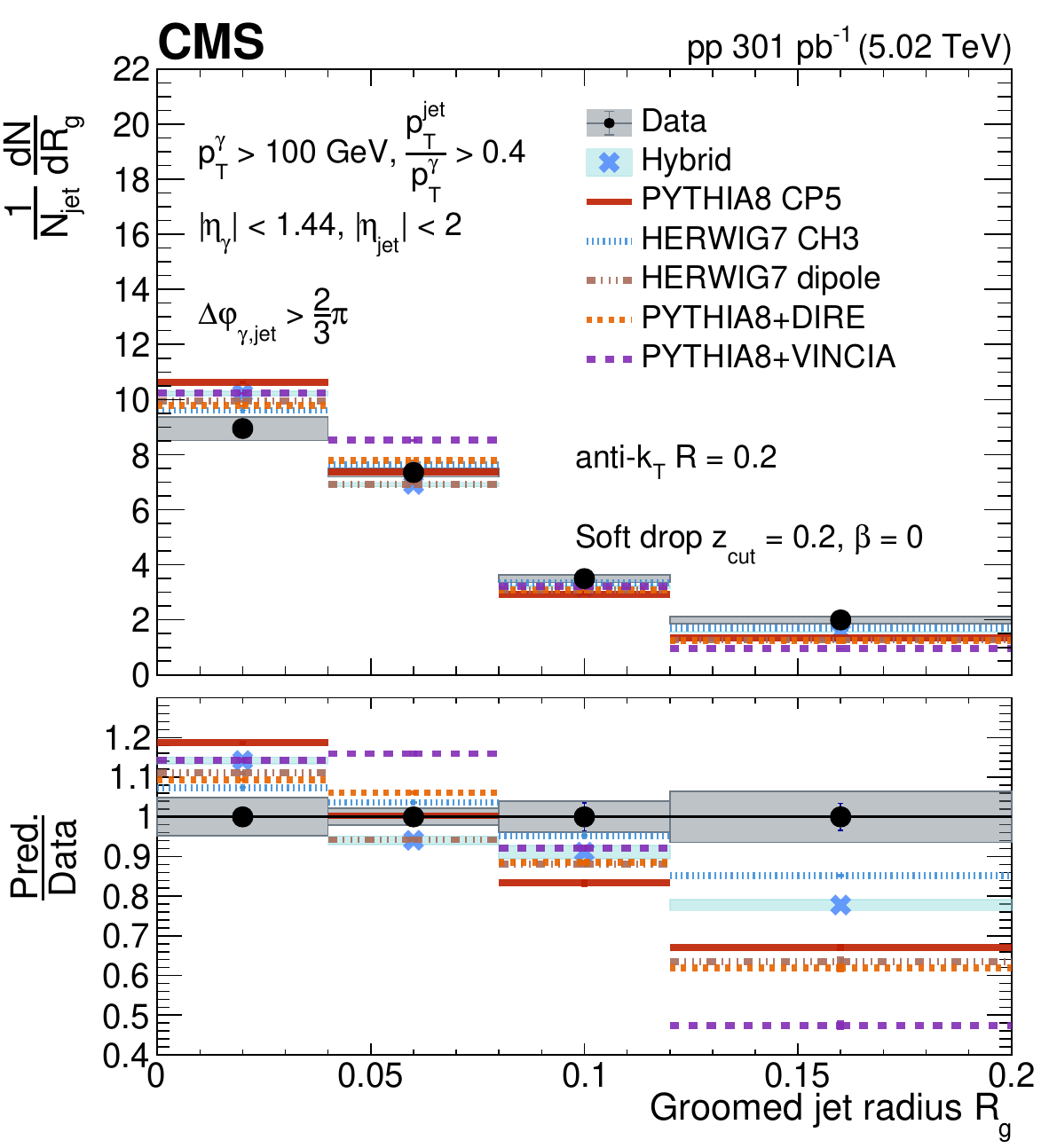}
    \caption{Unfolded distributions of jet girth $g$ (\cmsLeft) and groomed jet radius \Rg (\cmsRight) of photon-tagged jets in \pp collisions for $\xgJ \equiv \pTjet/\pTg > 0.4$. The upper panels show the comparison of the observable in \pp collisions and predictions from simulated events. The lower panels show the corresponding ratios of the MC calculations and data. The bands represent the total uncertainties, whereas the vertical bars represent the statistical uncertainties.}
    \label{fig:pp_vs_mcplot}
\end{figure}

\begin{figure}[htbp!]
    \centering
        \includegraphics[width=0.49\textwidth]{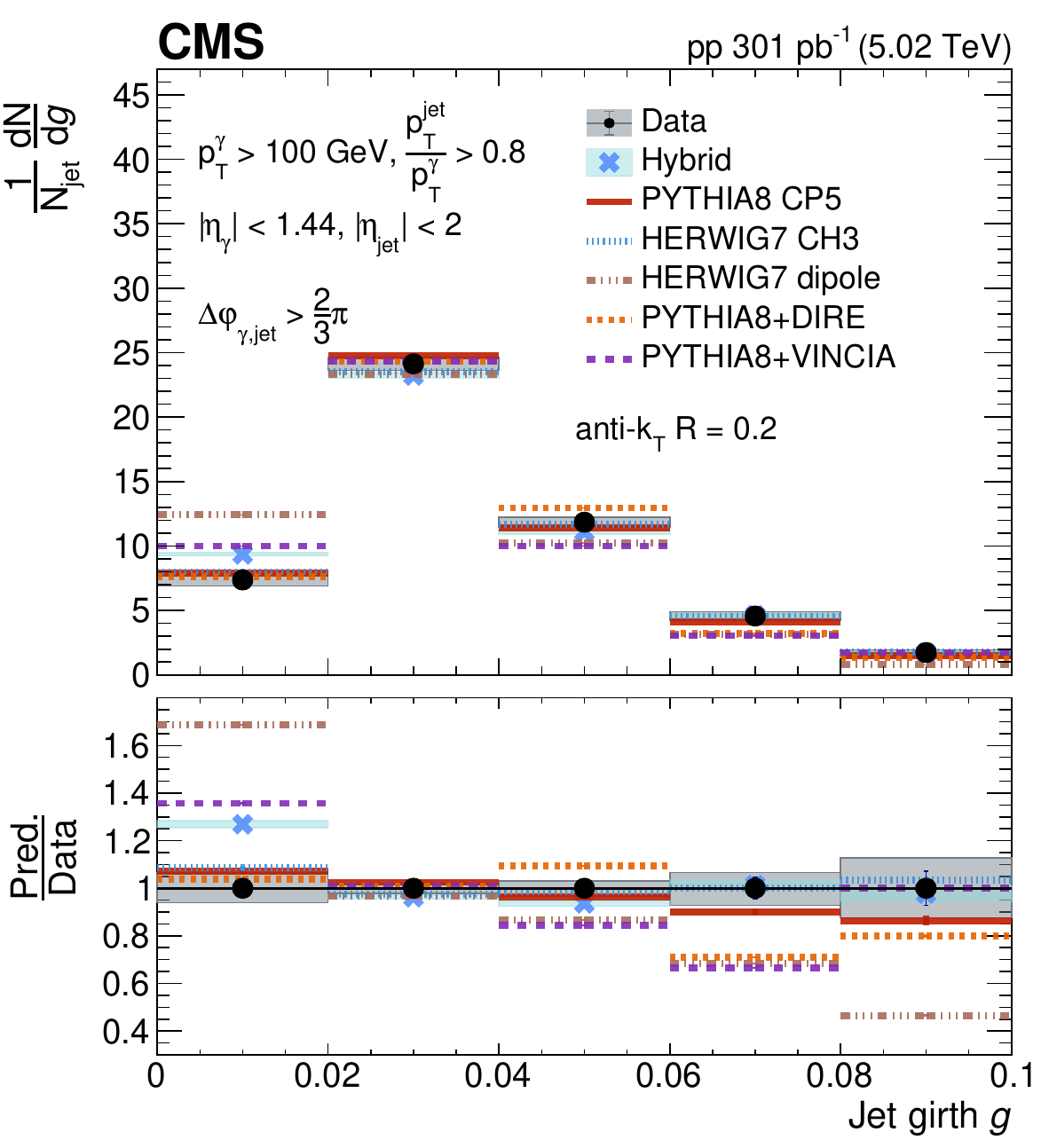}
        \includegraphics[width=0.49\textwidth]{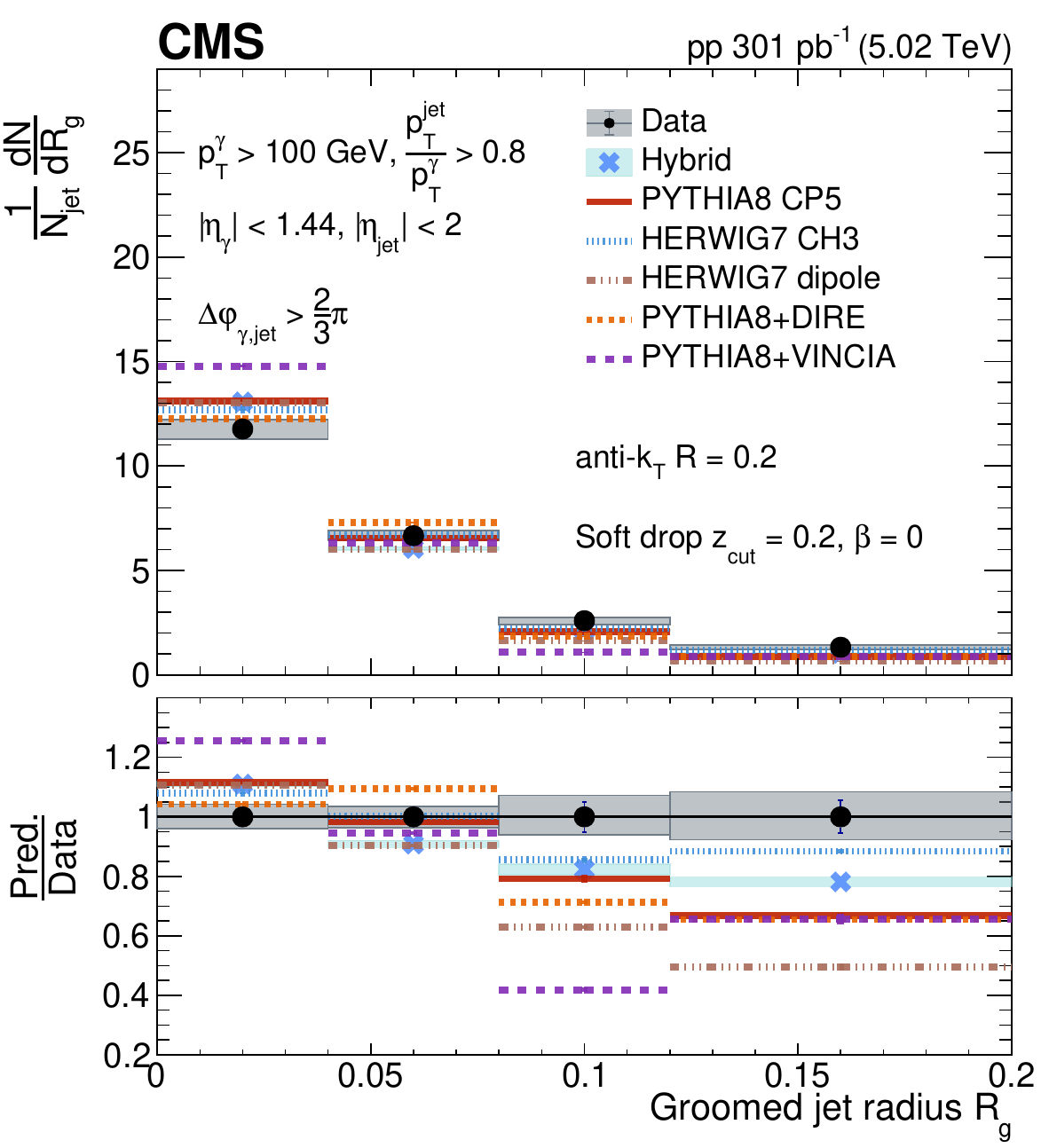}
    \caption{Unfolded distributions of jet girth $g$ (\cmsLeft) and groomed jet radius \Rg (\cmsRight) of photon-tagged jets in \pp collisions for $\xgJ \equiv \pTjet/\pTg > 0.8$. The upper panels show the comparison of the observable in \pp collisions and predictions from simulated events. The lower panels show the corresponding ratios of the MC calculations and data. The bands represent the total experimental uncertainties, whereas the vertical bars represent the statistical uncertainties.}
    \label{fig:pp_vs_mcplot_xJ_gp8}
\end{figure}

The unfolded distributions of $g$ and \Rg of photon-tagged jets in \pp collisions are shown in Figs.~\ref{fig:pp_vs_mcplot} and~\ref{fig:pp_vs_mcplot_xJ_gp8} for $\xgJ > 0.4$ and $\xgJ > 0.8$, respectively. In \pp collisions, one can use the $\xgJ > 0.4$ and 0.8 categories to better understand out-of-cone radiation effects, which lead to an average shift in jet \pt~\cite{Dasgupta:2007wa}. On average, gluon-initiated jets experience a larger \pt shift from out-of-cone radiation than quark-initiated jets~\cite{Dasgupta:2007wa}, so selecting events based on $\xgJ > 0.4$ and 0.8 leads to different parton flavor compositions.

The \pp measurement is compared with MC simulation results from \PYTHIA{ 8.230} CP5, \PYTHIA{ 8.303} with the \VINCIA shower~\cite{Giele:2007di}, \PYTHIA{ 8.303} with the \DIRE shower~\cite{Hoche:2015sya}, \HERWIG{ 7.2.2} with an angular-ordered shower using tune CH3, and \HERWIG{ 7.2.0} with a dipole shower~\cite{Bellm:2015jjp,Bellm:2019zci}. The \DIRE shower implements a \pt-ordered color dipole shower, where the radiator--spectator particle pairs evolve simultaneously, and it includes higher-order corrections, such as triple-collinear or double-soft parton emissions. The version of \VINCIA in \PYTHIA{ 8.303} uses the antenna sector shower formalism~\cite{Gehrmann-DeRidder:2005btv, Giele:2007di,Brooks:2020upa}. The branching kernels, known as antenna functions, treat coherent sums of parton pairs without requiring a separation into radiators and spectators. Both \VINCIA and \DIRE have their respective set of tuning parameters in \PYTHIA{8}. The dipole parton shower available in \HERWIG{7} uses the Catani--Seymour dipole factorization formalism~\cite{Catani:1996vz}.  We also compare the ``hybrid'' model predictions for vacuum radiation with the data~\cite{Casalderrey-Solana:2014bpa}, which consists of \PYTHIA{8} generator predictions with the Monash tune as a baseline~\cite{Skands:2014pea}, but without multiple parton interactions, a slightly larger \pt cutoff for the final-state radiation, and with the PDF set NNPDF2.3 QCD+QED LO~\cite{Ball:2014uwa}. All the predictions presented here use matrix elements at LO in perturbation theory for the hard scattering.

Differences between \pp unfolded distributions and the MC simulated predictions are of the order of 20\% in the bulk of the distributions. The largest discrepancies are at small and large values of $g$ and at large values of \Rg. This indicates that the jets measured in data are, on average, broader than the jets in simulation. The \PYTHIA{8} CP5 predictions describe the $g$ distribution for both $\xgJ > 0.4$ and 0.8 selections within the uncertainties, but not \Rg. The hybrid model predictions for \pp have similar trends as \PYTHIA{8} CP5. The best global description is provided by the \HERWIG{7} CH3 generator for both $\xgJ > 0.4$ and 0.8 categories in the $g$ and \Rg variables, whereas the \HERWIG{7} generator with a dipole shower tends to have the largest differences with the data. Recent measurements of jet substructure in \Zjet and dijet events in \pp collisions~\cite{CMS:2021iwu} suggest that the jet substructure modeling depends on details of the quark and gluon jet composition of the sample. Photon+jet events are expected to have a larger quark jet fraction relative to inclusive jets for a similar jet \pt domain. Thus, the present measurement can help constrain the modeling of quark jet fragmentation in simulations, which can be useful to improve the modeling of the vacuum shower baseline used in predictions for medium-induced jet modification.

\begin{figure}[htbp!]
    \centering
        \includegraphics[width=0.49\textwidth]{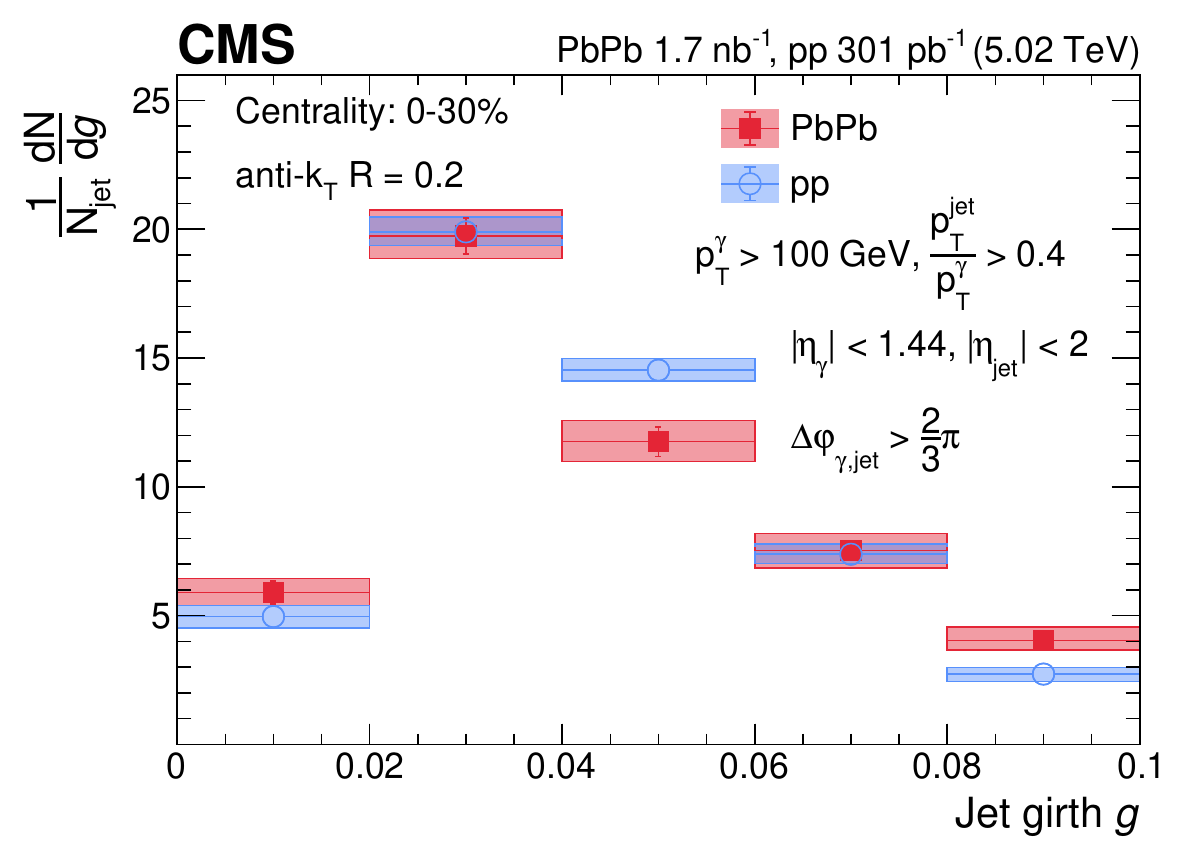}
        \includegraphics[width=0.49\textwidth]{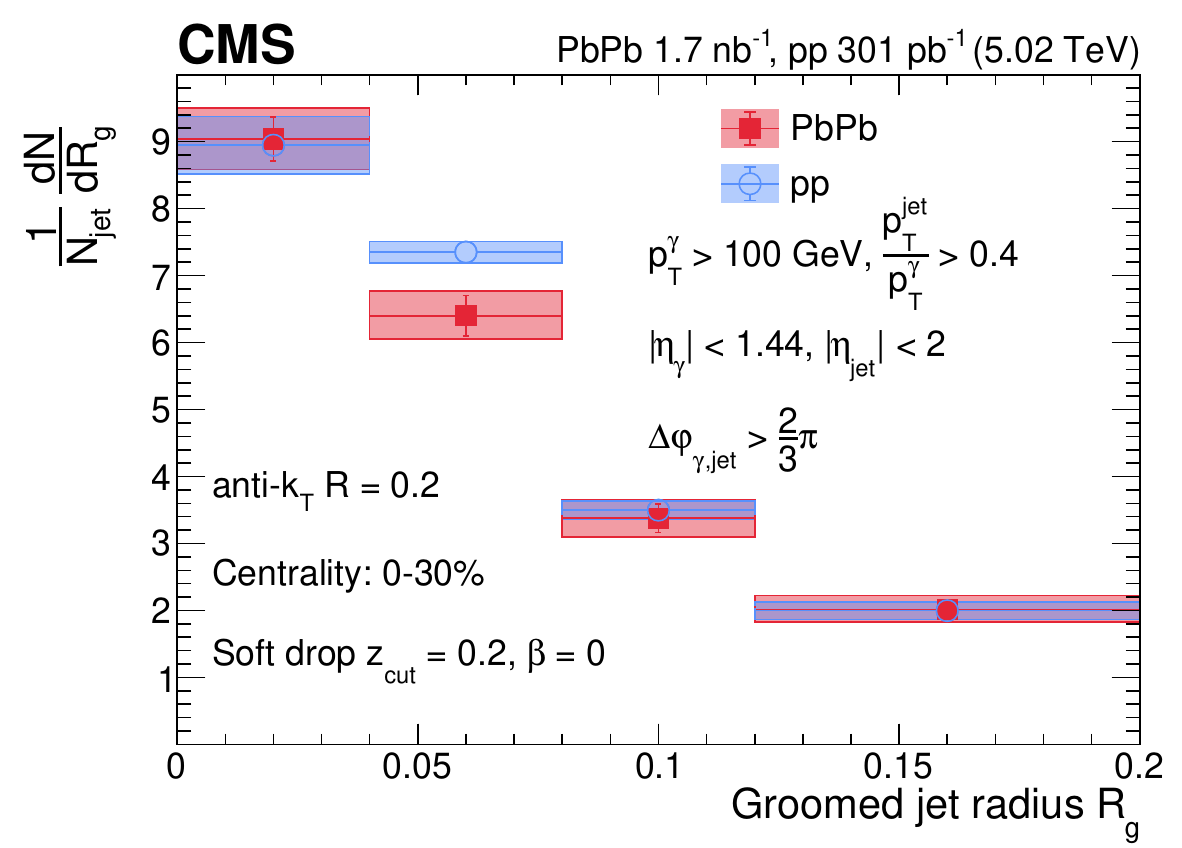}
    \caption{Unfolded distributions of jet girth $g$ (\cmsLeft) and groomed jet radius \Rg (\cmsRight) of photon-tagged jets in \PbPb and \pp collisions for $\xgJ \equiv \pTjet/\pTg > 0.4$ (selecting both more and less quenched jets). The bands represent the total uncertainties, whereas the vertical bars represent the statistical uncertainties.}
    \label{fig:Raw_unfolded_pp_PbPb_0_30_xJ_gp4}
\end{figure}

\begin{figure}[htbp!]
    \centering
        \includegraphics[width=0.49\textwidth]{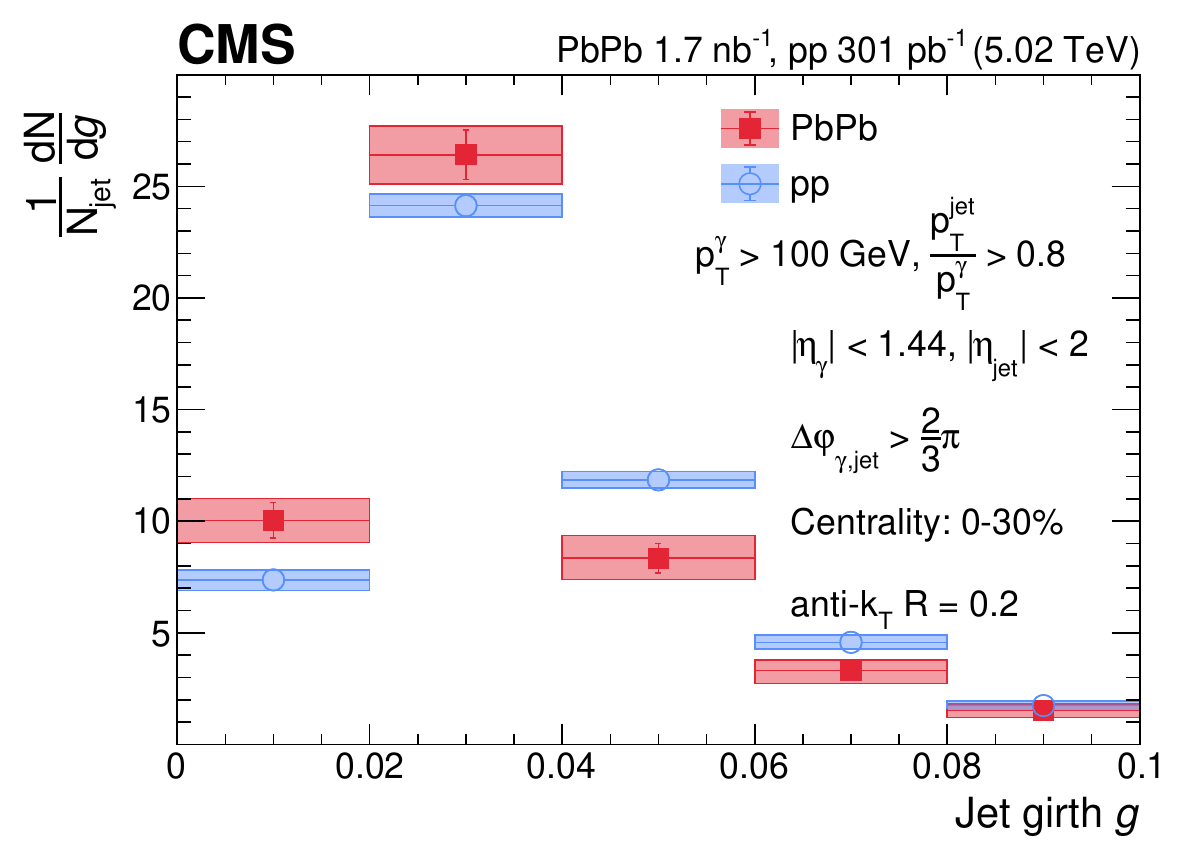}
        \includegraphics[width=0.49\textwidth]{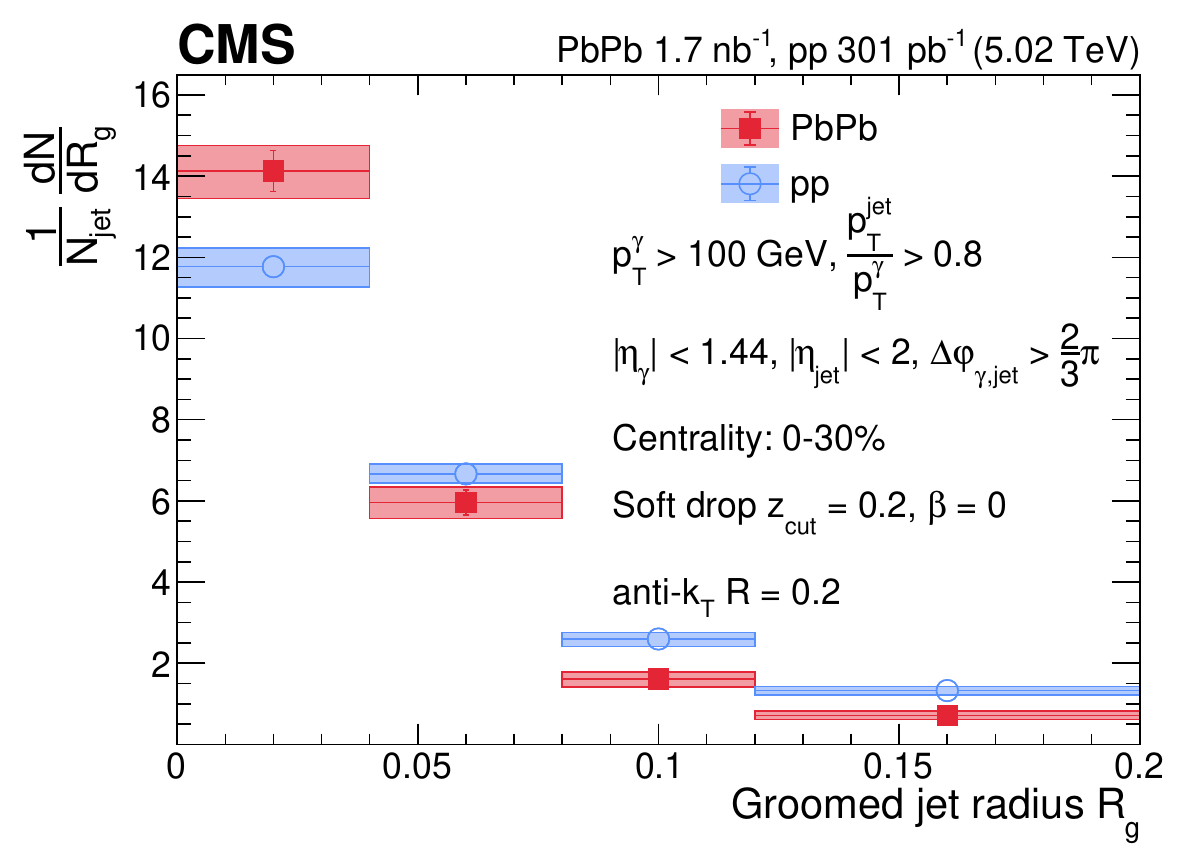}
        
    \caption{Unfolded distributions of jet girth $g$ (\cmsLeft) and groomed jet radius \Rg (\cmsRight) of photon-tagged jets in \PbPb and \pp collisions for $\xgJ \equiv \pTjet/\pTg > 0.8$ (selecting less quenched jets). The bands represent the total uncertainties, whereas the vertical bars represent the statistical uncertainties.}
    \label{fig:Raw_unfolded_pp_PbPb_0_30_xJ_gp8}
\end{figure}

\begin{figure*}[pt!]
    \centering
        \includegraphics[width=0.49\textwidth]{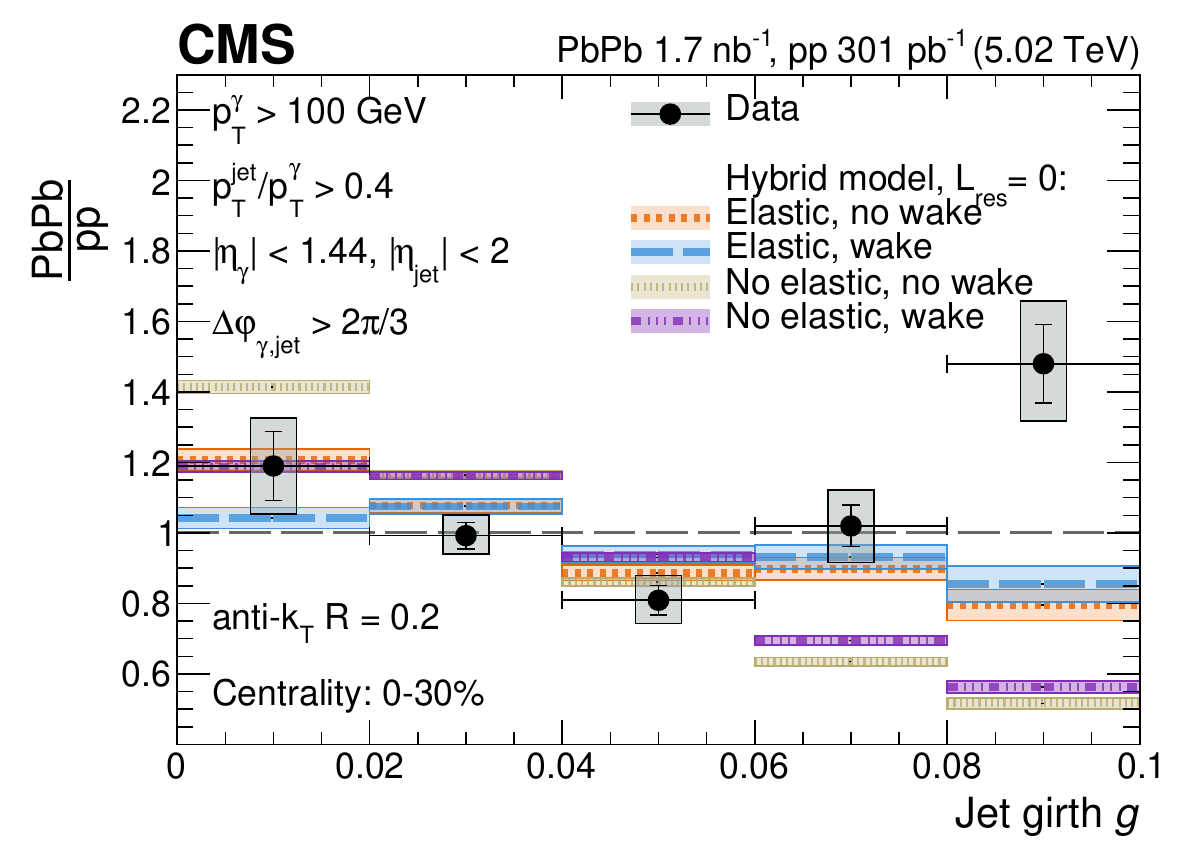}
        \includegraphics[width=0.49\textwidth]{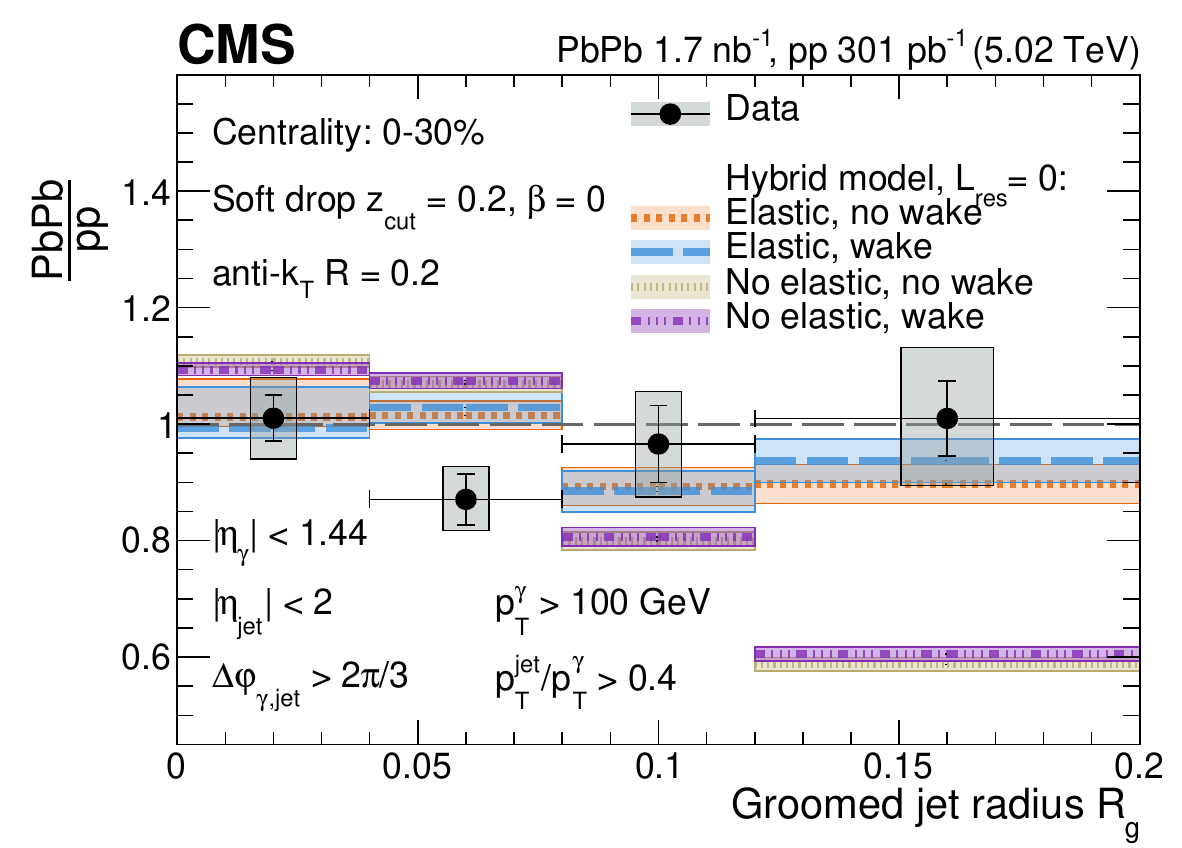} \\
        \includegraphics[width=0.49\textwidth]{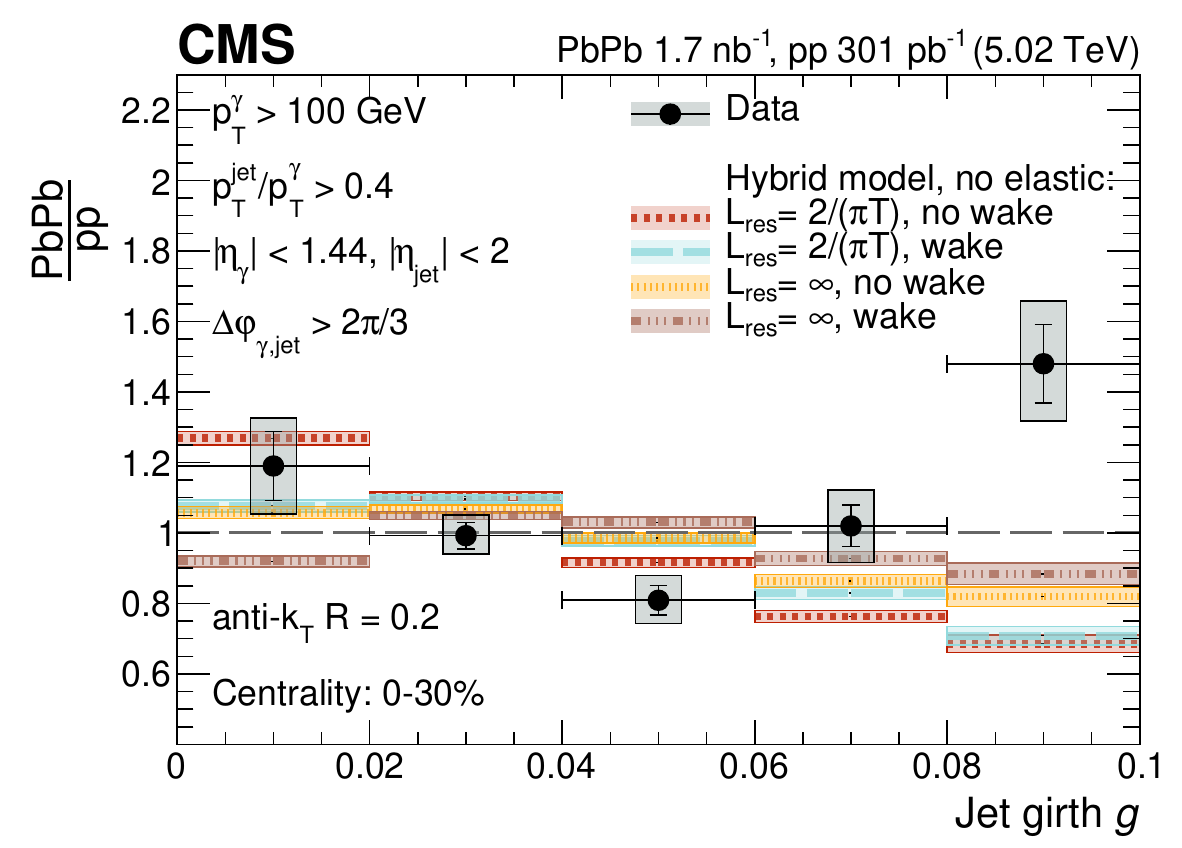}
        \includegraphics[width=0.49\textwidth]{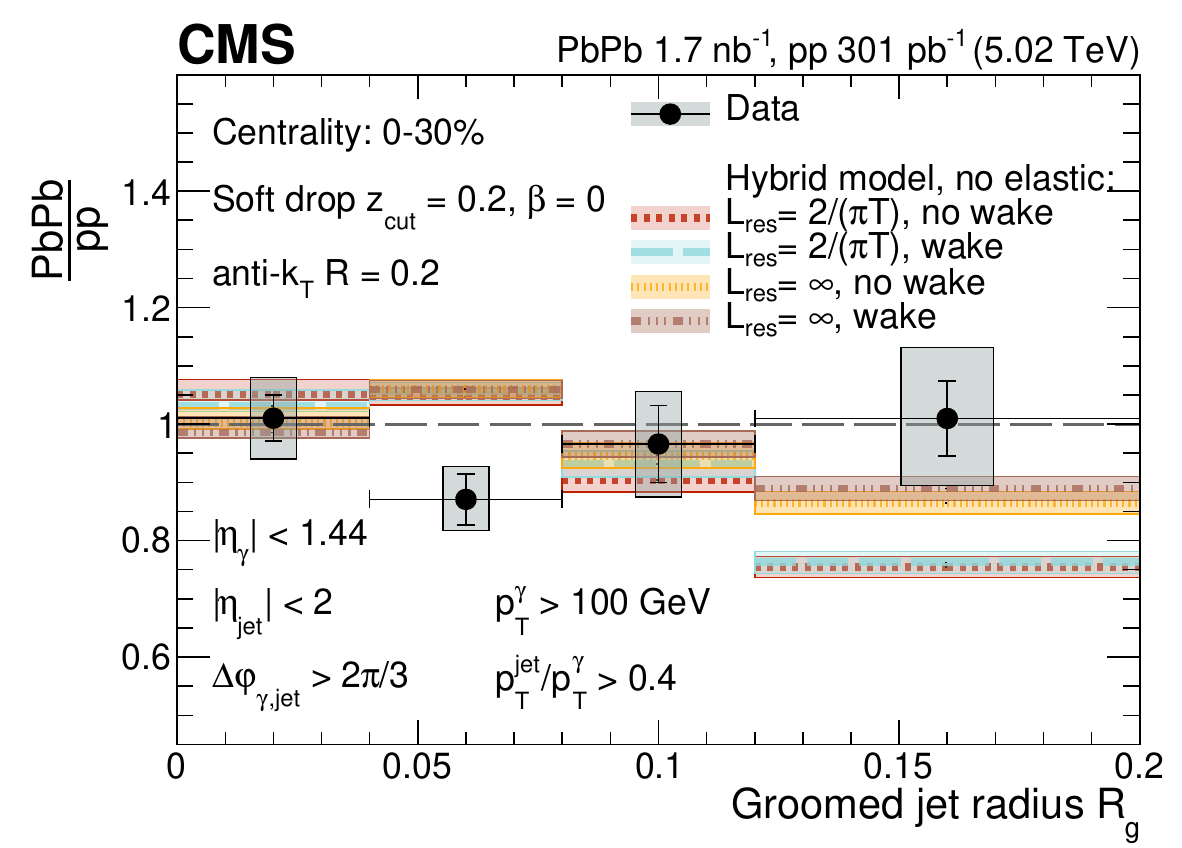}
        
    \caption{Ratio of the normalized yields of \PbPb to \pp data for jet girth $g$ (left) and groomed jet radius \Rg (right) of photon-tagged jets in \PbPb and \pp collisions for $\xgJ \equiv \pTjet/\pTg > 0.4$ (selecting both more and less quenched jets). The data are compared with the hybrid model predictions for $\Lres = 0$ (upper) and for nonzero values of \Lres without elastic scattering (lower). The bands around the data points represent the total experimental uncertainties, whereas the vertical bars represent the statistical uncertainties. The uncertainties in the \PbPb-to-\pp ratio have been obtained assuming the \PbPb and \pp measurements are uncorrelated. The bands around the theory predictions represent the statistical uncertainties of the prediction.}
    \label{fig:Ratio_pp_PbPb_0_30_xJ_gp4}
\end{figure*}

\begin{figure*}[pt!]
    \centering
        \includegraphics[width=0.49\textwidth]{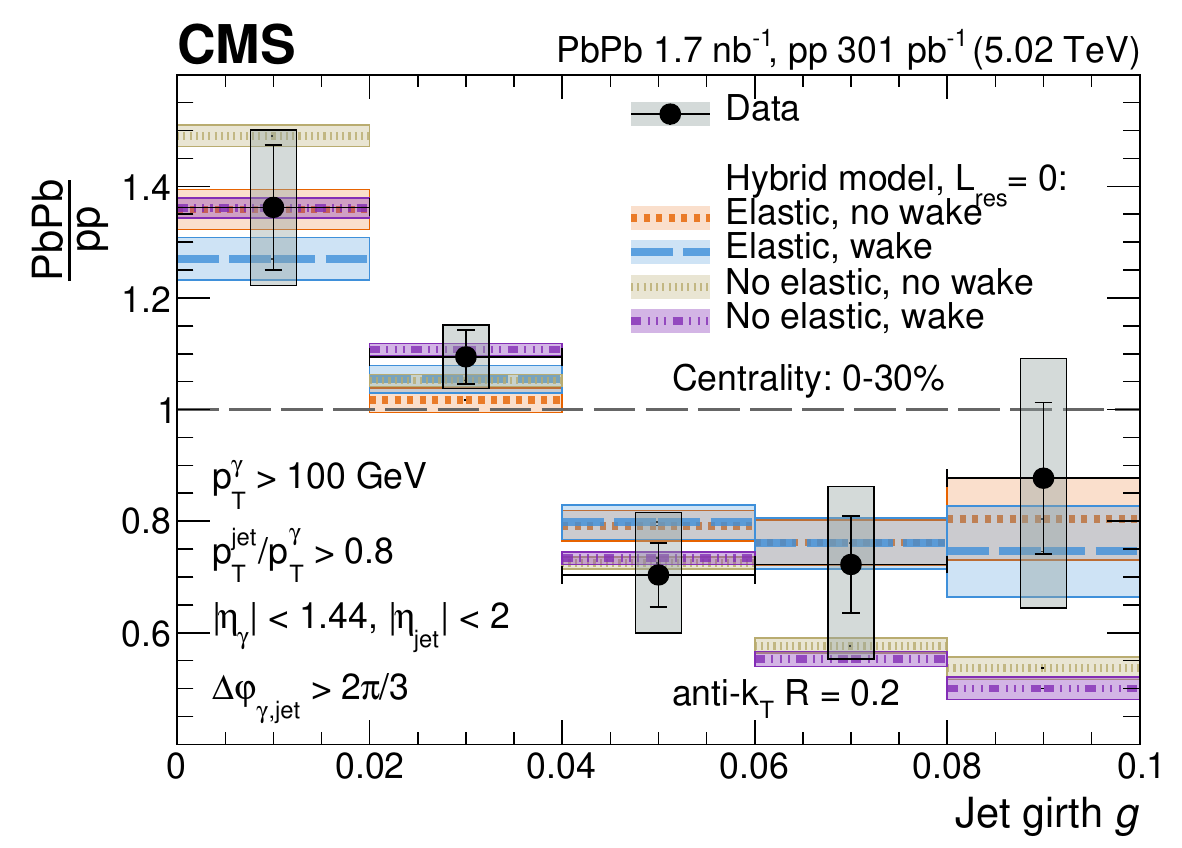}
        \includegraphics[width=0.49\textwidth]{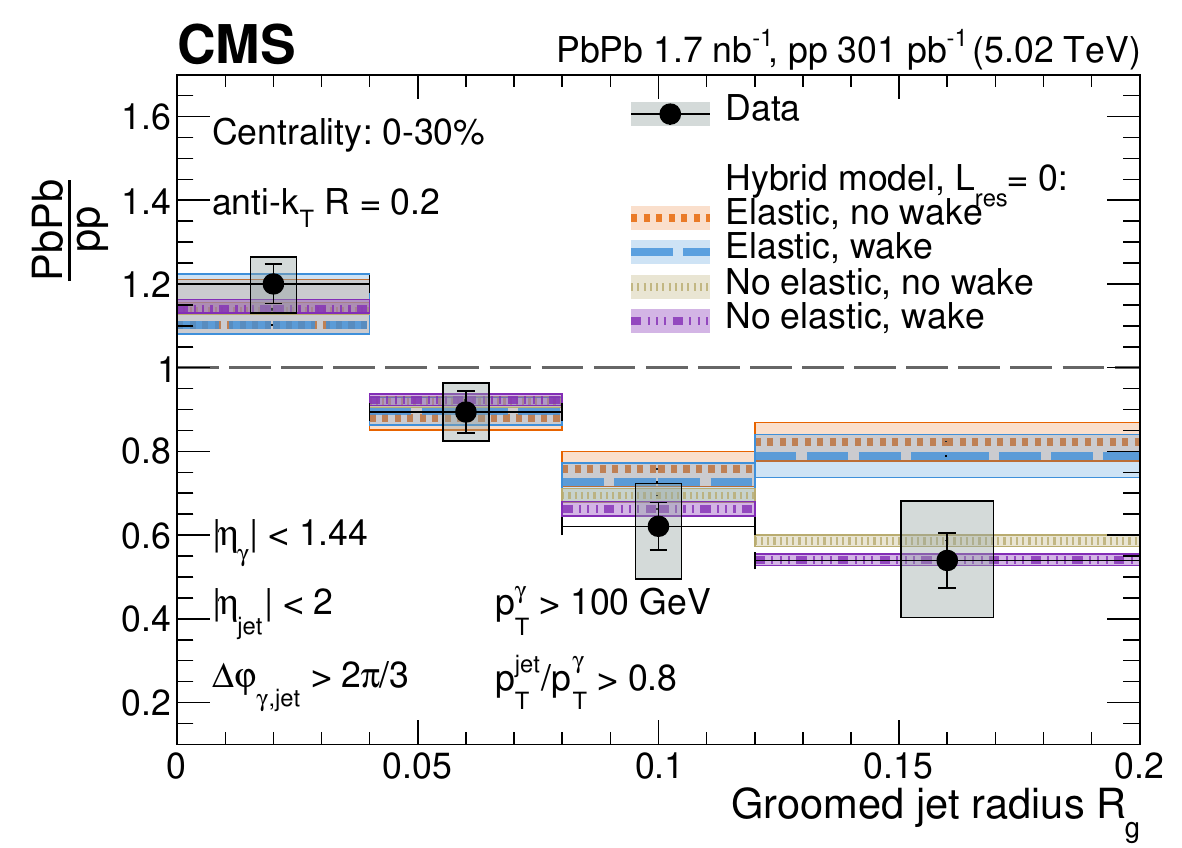} \\
        \includegraphics[width=0.49\textwidth]{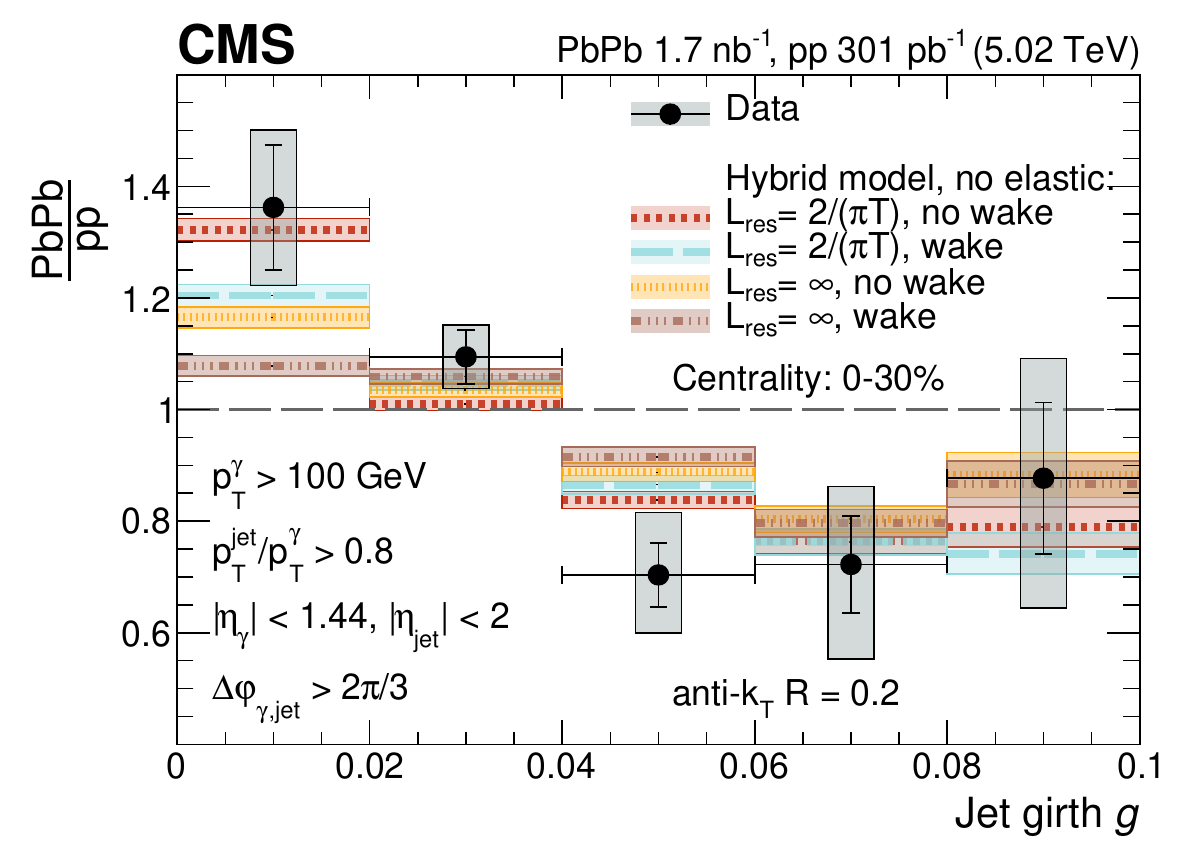}
        \includegraphics[width=0.49\textwidth]{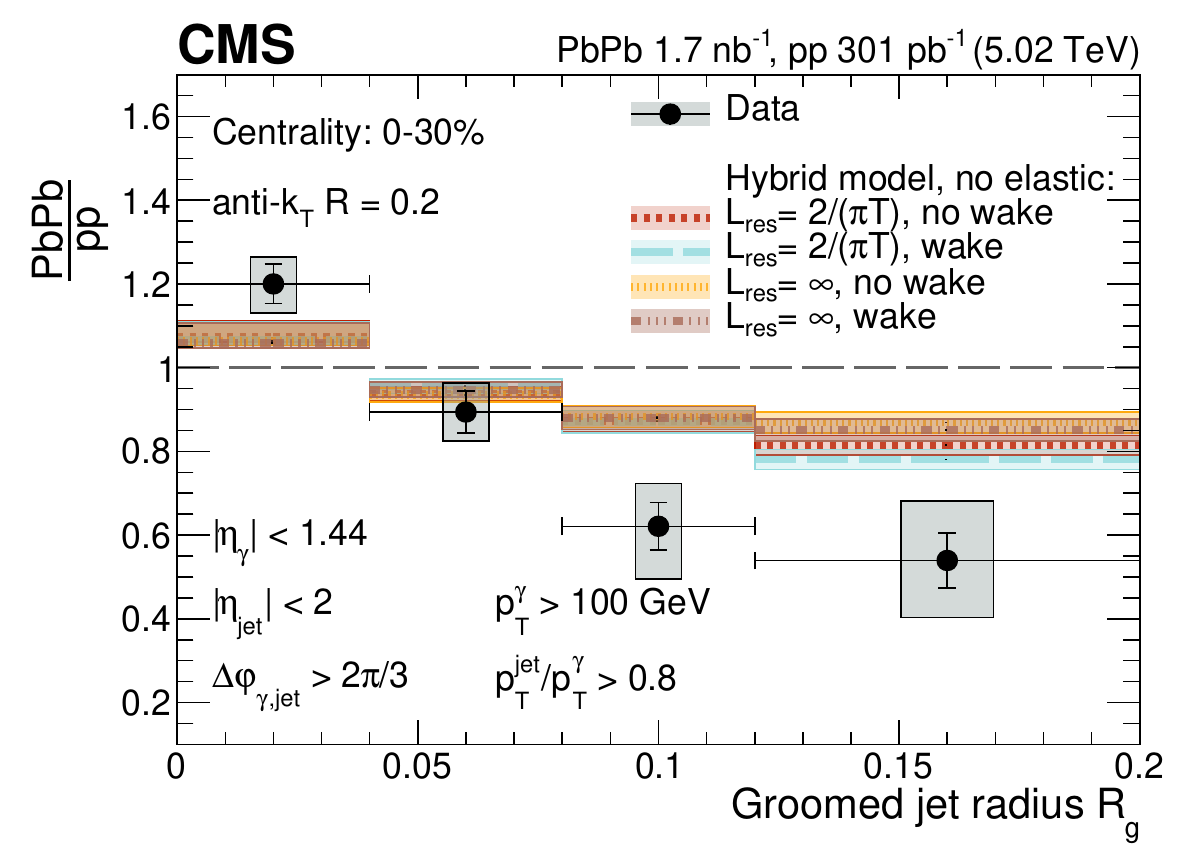}
        
    \caption{Ratio of the normalized yields of \PbPb to \pp for jet girth $g$ (left) and groomed jet radius \Rg (right) of photon-tagged jets in \PbPb and \pp collisions for $\xgJ \equiv \pTjet/\pTg > 0.8$ (selecting less quenched jets). The data are compared with the hybrid predictions for $\Lres = 0$ (upper) and nonzero values of \Lres without elastic scatterings (lower). The bands around the data points represent the total experimental uncertainties, whereas the vertical bars represent the statistical uncertainties. The uncertainties in the \PbPb-to-\pp ratio have been obtained assuming the \PbPb and \pp measurements are uncorrelated. The bands around the theory predictions represent the statistical uncertainties of the prediction.}
    \label{fig:Ratio_pp_PbPb_0_30_xJ_gp8}
\end{figure*}

The unfolded distributions of $g$ and \Rg for photon-tagged jets in \PbPb collisions compared with those of \pp collisions are shown in Figs.~\ref{fig:Raw_unfolded_pp_PbPb_0_30_xJ_gp4} and~\ref{fig:Raw_unfolded_pp_PbPb_0_30_xJ_gp8} for $\xgJ > 0.4$ and 0.8. The ratios of the unfolded distributions of \PbPb to \pp are shown in Fig.~\ref{fig:Ratio_pp_PbPb_0_30_xJ_gp4} for $\xgJ > 0.4$ and in Fig.~\ref{fig:Ratio_pp_PbPb_0_30_xJ_gp8} for $\xgJ > 0.8$. The ratios of \PbPb to \pp distributions are used to identify potential medium-induced modifications of the jet shower. For photon-tagged jets with $\xgJ > 0.4$ in Fig.~\ref{fig:Raw_unfolded_pp_PbPb_0_30_xJ_gp4}, which corresponds to a more inclusive jet selection where both quenched and less quenched jets are selected, we observe no narrowing of the angular structure of jets produced in \PbPb collisions relative to the distributions in \pp collisions within the experimental uncertainties. For $g$, there are hints of a broadening of the substructure of the jet in \PbPb collisions at large values of $g$, whereas \Rg is consistent with \pp within the experimental uncertainties. On the other hand, by selecting less quenched jets via $\xgJ > 0.8$ in Fig.~\ref{fig:Raw_unfolded_pp_PbPb_0_30_xJ_gp8}, which is a less inclusive jet selection, we observe a narrowing of the angular structure of jets in \PbPb collisions when compared with \pp collisions, contrary to the trend observed in events selected with $\xgJ > 0.4$. Figures ~\ref{fig:Ratio_pp_PbPb_0_30_xJ_gp4} and~\ref{fig:Ratio_pp_PbPb_0_30_xJ_gp8} compare the measurements of the ratio of the unfolded distributions of \PbPb to \pp with the hybrid model~\cite{Casalderrey-Solana:2014bpa}. The model is a ``hybrid'' approach of weak and strong coupling approximations to describe medium-induced jet modifications. It includes the modeling of energy loss due to the strong coupling between the partons and the medium, large-angle deflections of partons transversing the QGP, referred to as ``Moli\'{e}re elastic scatterings''~\cite{DEramo:2012uzl,DEramo:2018eoy}, and the nonperturbative backreaction of the medium~\cite{Casalderrey-Solana:2020rsj}, also known as medium response. These effects are denoted as ``elastic'' and ``wake'' in Figs.~\ref{fig:Ratio_pp_PbPb_0_30_xJ_gp4} and~\ref{fig:Ratio_pp_PbPb_0_30_xJ_gp8}, and the absence of such effects as ``no elastic'' or ``no wake.'' In the model, the medium resolution length is controlled by the parameter \Lres~\cite{Hulcher:2017cpt}, which corresponds to the minimum transverse length between two color-connected partons for the medium to resolve them separately. Three values of \Lres are considered: $\Lres = 0$, $2/(\pi T),$ and $  \infty$, where $T$ is the temperature of the medium in the model. The $\Lres = 0$ value corresponds to the incoherence limit where all the radiators are resolved by the medium. The $\Lres = 2/(\pi T)$ corresponds to the expectation that \Lres should be of the same order of magnitude as the Debye screening length~\cite{Hulcher:2017cpt} and it represents an intermediate scenario where only a fraction of the radiators interact with the medium. The  $\Lres = \infty$ value corresponds to the full coherence limit, where radiators are not resolved individually and the jet interacts with the medium as a single color charge. In the case of elastic scatterings with the medium, the present model accounts for them for $\Lres = 0$.

Some conclusions can be drawn from the comparison of these theory predictions from the hybrid model to the experimental data. First, the contribution of the wake is negligible, as shown in Figs.~\ref{fig:Ratio_pp_PbPb_0_30_xJ_gp4} and~\ref{fig:Ratio_pp_PbPb_0_30_xJ_gp8}, because of the small $R$ parameter used in the measurement. Second, there is no consistent choice of parameters that can describe both the $\xgJ > 0.4$ and  0.8 categories simultaneously. For $\xgJ > 0.4$ in Fig.~\ref{fig:Ratio_pp_PbPb_0_30_xJ_gp4}, the incoherence case $\Lres = 0$ overestimates the narrowing, while the model describes the data better when the Moli{\`e}re elastic scatterings are added, which broaden the jet, or when the limiting case of $\Lres = \infty$ is considered. The $\xgJ > 0.4$ data sample alone cannot be used to separate the contribution of color coherence and elastic scatterings, but when used together with the unfolded distributions of $\xgJ > 0.8$, one can better separate the consequences of such effects, as shown in Fig.~\ref{fig:Ratio_pp_PbPb_0_30_xJ_gp8}. The incoherence limit $\Lres = 0$ is favored in the $\xgJ > 0.8$ category, and both the effects of elastic scatterings and the effect of nonzero \Lres values result in a slight overestimation of the normalized yield at large \Rg values. For $g$ in Figs.~\ref{fig:Ratio_pp_PbPb_0_30_xJ_gp4} and~\ref{fig:Ratio_pp_PbPb_0_30_xJ_gp8}, the results of the model comparison to the data are qualitatively the same as for \Rg, with an underestimation of the \PbPb-to-\pp ratio at large values of $g$ from all setups for $\xgJ > 0.4$. Since the categories $\xgJ > 0.4$ and 0.8 enable the selection of different degrees of jet energy loss, the present measurement has sensitivity on how the different physical mechanisms affect the radiation pattern of the jet as well as the jet energy loss. Thus, the present measurement can be used to optimize the model parameter choice together with existing measurements that are sensitive to other aspects of medium-induced jet modifications.

The measurement using photon-tagged jets is complementary to previous measurements in inclusive jet events. In inclusive jet measurements, the comparison between jets in \PbPb and \pp collisions is done at the same reconstructed jet \pt, but due to the jet energy loss that occurs in \PbPb collisions, the comparison is not done for the same initial parton energies. In contrast to the trends observed by the ALICE and ATLAS Collaborations for \Rg or $g$ in inclusive jet events~\cite{ALargeIonColliderExperiment:2021mqf,ATLAS:2022vii}, we do not observe a narrowing of the substructure of jets in \Rg or $g$ within the experimental uncertainties when selecting jets with $\xgJ > 0.4$ and $\pTg > 100\GeV$. One possible explanation is that, after reducing the selection bias effect due to jet energy loss, the sample of events being analyzed in photon-tagged events consists not only of narrow unquenched jets but also of broader, and thus more quenched, jets. By selecting less quenched jets via $\xgJ > 0.8$, we observe an effective narrowing in the angular substructure of jets produced in \PbPb collisions, qualitatively similar to the distributions measured by the ALICE and ATLAS Collaborations~\cite{ALargeIonColliderExperiment:2021mqf,ATLAS:2022vii}. We verified that the conclusions of the study are the same even after imposing a requirement on $\xgJ < 1$ to reduce the contribution of bremsstrahlung photons.  It should be noted that since quark jets are on average harder and have narrower radiation patterns than gluon jets in a vacuum, a substructure-dependent energy loss --as expected from jet quenching theoretical calculations that rely on the dominance of the early-vacuum shower-- would be naturally correlated with parton flavor. 

\section{Summary}
\label{sec:summary}

In summary, we report the first measurements girth $g$ and the groomed jet radius \Rg of jets recoiling against isolated photons in lead-lead (\PbPb) and proton-proton (\pp) collisions. The analysis uses \PbPb and \pp collision data, both at a nucleon-nucleon center-of-mass energy of 5.02\TeV. The distributions are unfolded to the particle level in order to facilitate comparisons between experiments and with theoretical predictions.

The transverse momentum \pt of isolated photons ($\pTg$) can be used as a proxy for the \pt of the high-virtuality parton that initiates the shower of the recoiling jet. This enables the disentanglement of the potential modification of the momentum and angular substructure of jets due to the interactions with the medium from the selection bias effects that can originate from jet energy loss. This is done using the transverse momentum imbalance, defined as the ratio of the hardest recoil jet \pt (\pTjet) and \pTg, $\xgJ \equiv \pTjet/\pTg$. It is found that jets with $\pTjet/\pTg > 0.8$, \ie, those that closely balance the photon $\pTg$, are narrower in \PbPb than in \pp collisions. Relaxing the selection to include jets with $\pTjet/\pTg > 0.4$ reduces the narrowing of the angular structure of jets in \PbPb relative to the \pp reference. These observations suggest that selection bias effects play an important role in the interpretation of the modification of the angular scales of jets in terms of medium-induced effects. The measured distributions are compared with calculations based on a hybrid strong and weak coupling model to describe medium-induced jet modifications. According to model predictions, the \Rg and $g$ distributions are not very sensitive to medium response effects or to variations of the medium resolution length. However, changes in the modeling of Moli{\`e}re elastic scatterings have an effect of 10--40\% at large values of $g$ and \Rg. This shows the ability of the data to constrain the impact of Moli{\`e}re scatterings in a way that is effectively factorized from the effects of the wake and the medium resolution length.

Medium-induced jet modifications are commonly assessed by comparing jets and their substructure at the same reconstructed \pt in \PbPb and \pp collisions, which in the former case corresponds to the momentum of the jet after its interactions with the quark gluon plasma. These interactions are expected to broaden the jet and reduce its energy. Thus, in an inclusive jet measurement, when comparing populations of jets in \PbPb and pp within the same measured jet \pt window, a selection bias can lead to an effective narrowing of the angular structure of jets in \PbPb relative to \pp. One possibility is that the population of jets that were initially broader (hence, more strongly quenched jets) has migrated to lower jet energies, whereas the population of narrower jets (less strongly quenched jets) remains. Thus, events with high-\pt jets recoiling against energetic isolated photons can be used to better constrain genuine medium modifications of the jet shower, complementing measurements in inclusive jet production.

\begin{acknowledgments}
We thank Daniel Pablos and Krishna Rajagopal for providing their theoretical calculations of the groomed jet radius and jet girth adapted with the selection requirements used in this analysis.

We congratulate our colleagues in the CERN accelerator departments for the excellent performance of the LHC and thank the technical and administrative staffs at CERN and at other CMS institutes for their contributions to the success of the CMS effort. In addition, we gratefully acknowledge the computing centers and personnel of the Worldwide LHC Computing Grid and other centers for delivering so effectively the computing infrastructure essential to our analyses. Finally, we acknowledge the enduring support for the construction and operation of the LHC, the CMS detector, and the supporting computing infrastructure provided by the following funding agencies: SC (Armenia), BMBWF and FWF (Austria); FNRS and FWO (Belgium); CNPq, CAPES, FAPERJ, FAPERGS, and FAPESP (Brazil); MES and BNSF (Bulgaria); CERN; CAS, MoST, and NSFC (China); MINCIENCIAS (Colombia); MSES and CSF (Croatia); RIF (Cyprus); SENESCYT (Ecuador); ERC PRG, RVTT3 and MoER TK202 (Estonia); Academy of Finland, MEC, and HIP (Finland); CEA and CNRS/IN2P3 (France); SRNSF (Georgia); BMBF, DFG, and HGF (Germany); GSRI (Greece); NKFIH (Hungary); DAE and DST (India); IPM (Iran); SFI (Ireland); INFN (Italy); MSIP and NRF (Republic of Korea); MES (Latvia); LMTLT (Lithuania); MOE and UM (Malaysia); BUAP, CINVESTAV, CONACYT, LNS, SEP, and UASLP-FAI (Mexico); MOS (Montenegro); MBIE (New Zealand); PAEC (Pakistan); MES and NSC (Poland); FCT (Portugal);  MESTD (Serbia); MCIN/AEI and PCTI (Spain); MOSTR (Sri Lanka); Swiss Funding Agencies (Switzerland); MST (Taipei); MHESI and NSTDA (Thailand); TUBITAK and TENMAK (Turkey); NASU (Ukraine); STFC (United Kingdom); DOE and NSF (USA).

\hyphenation{Rachada-pisek} Individuals have received support from the Marie-Curie program and the European Research Council and Horizon 2020 Grant, contract Nos.\ 675440, 724704, 752730, 758316, 765710, 824093, 101115353, 101002207, and COST Action CA16108 (European Union); the Leventis Foundation; the Alfred P.\ Sloan Foundation; the Alexander von Humboldt Foundation; the Science Committee, project no. 22rl-037 (Armenia); the Belgian Federal Science Policy Office; the Fonds pour la Formation \`a la Recherche dans l'Industrie et dans l'Agriculture (FRIA-Belgium); the Agentschap voor Innovatie door Wetenschap en Technologie (IWT-Belgium); the F.R.S.-FNRS and FWO (Belgium) under the ``Excellence of Science -- EOS" -- be.h project n.\ 30820817; the Beijing Municipal Science \& Technology Commission, No. Z191100007219010 and Fundamental Research Funds for the Central Universities (China); the Ministry of Education, Youth and Sports (MEYS) of the Czech Republic; the Shota Rustaveli National Science Foundation, grant FR-22-985 (Georgia); the Deutsche Forschungsgemeinschaft (DFG), under Germany's Excellence Strategy -- EXC 2121 ``Quantum Universe" -- 390833306, and under project number 400140256 - GRK2497; the Hellenic Foundation for Research and Innovation (HFRI), Project Number 2288 (Greece); the Hungarian Academy of Sciences, the New National Excellence Program - \'UNKP, the NKFIH research grants K 131991, K 133046, K 138136, K 143460, K 143477, K 146913, K 146914, K 147048, 2020-2.2.1-ED-2021-00181, and TKP2021-NKTA-64 (Hungary); the Council of Science and Industrial Research, India; ICSC -- National Research Center for High Performance Computing, Big Data and Quantum Computing and FAIR -- Future Artificial Intelligence Research, funded by the NextGenerationEU program (Italy); the Latvian Council of Science; the Ministry of Education and Science, project no. 2022/WK/14, and the National Science Center, contracts Opus 2021/41/B/ST2/01369 and 2021/43/B/ST2/01552 (Poland); the Funda\c{c}\~ao para a Ci\^encia e a Tecnologia, grant CEECIND/01334/2018 (Portugal); the National Priorities Research Program by Qatar National Research Fund; MCIN/AEI/10.13039/501100011033, ERDF ``a way of making Europe", and the Programa Estatal de Fomento de la Investigaci{\'o}n Cient{\'i}fica y T{\'e}cnica de Excelencia Mar\'{\i}a de Maeztu, grant MDM-2017-0765 and Programa Severo Ochoa del Principado de Asturias (Spain); the Chulalongkorn Academic into Its 2nd Century Project Advancement Project, and the National Science, Research and Innovation Fund via the Program Management Unit for Human Resources \& Institutional Development, Research and Innovation, grant B37G660013 (Thailand); the Kavli Foundation; the Nvidia Corporation; the SuperMicro Corporation; the Welch Foundation, contract C-1845; and the Weston Havens Foundation (USA).

\end{acknowledgments}

\bibliography{auto_generated}\cleardoublepage \appendix\section{The CMS Collaboration \label{app:collab}}\begin{sloppypar}\hyphenpenalty=5000\widowpenalty=500\clubpenalty=5000
\cmsinstitute{Yerevan Physics Institute, Yerevan, Armenia}
{\tolerance=6000
A.~Hayrapetyan, A.~Tumasyan\cmsAuthorMark{1}\cmsorcid{0009-0000-0684-6742}
\par}
\cmsinstitute{Institut f\"{u}r Hochenergiephysik, Vienna, Austria}
{\tolerance=6000
W.~Adam\cmsorcid{0000-0001-9099-4341}, J.W.~Andrejkovic, T.~Bergauer\cmsorcid{0000-0002-5786-0293}, S.~Chatterjee\cmsorcid{0000-0003-2660-0349}, K.~Damanakis\cmsorcid{0000-0001-5389-2872}, M.~Dragicevic\cmsorcid{0000-0003-1967-6783}, P.S.~Hussain\cmsorcid{0000-0002-4825-5278}, M.~Jeitler\cmsAuthorMark{2}\cmsorcid{0000-0002-5141-9560}, N.~Krammer\cmsorcid{0000-0002-0548-0985}, A.~Li\cmsorcid{0000-0002-4547-116X}, D.~Liko\cmsorcid{0000-0002-3380-473X}, I.~Mikulec\cmsorcid{0000-0003-0385-2746}, J.~Schieck\cmsAuthorMark{2}\cmsorcid{0000-0002-1058-8093}, R.~Sch\"{o}fbeck\cmsorcid{0000-0002-2332-8784}, D.~Schwarz\cmsorcid{0000-0002-3821-7331}, M.~Sonawane\cmsorcid{0000-0003-0510-7010}, S.~Templ\cmsorcid{0000-0003-3137-5692}, W.~Waltenberger\cmsorcid{0000-0002-6215-7228}, C.-E.~Wulz\cmsAuthorMark{2}\cmsorcid{0000-0001-9226-5812}
% ...
\par}
\vskip\cmsinstskip
\dag:~Deceased\\
$^{1}$Also at Yerevan State University, Yerevan, Armenia\\
$^{2}$Also at TU Wien, Vienna, Austria\\
% ...
\end{sloppypar}
%%% END EDITABLE REGION %%%
% skeleton_end
\end{document}